\documentclass[preprint,eqsecnum,preprintnumbers,nofootinbib,byrevtex,prd,aps,showpacs,showkeys,groupedaddress,floatfix]{revtex4}
\usepackage{bm}
\usepackage{graphics}
\usepackage{graphicx}
\usepackage{epsfig}
\usepackage{amssymb}
\usepackage{amsmath}

\newcommand\nn{\nonumber}
\newcommand\ba{\begin{eqnarray}}
\newcommand\ea{\end{eqnarray}}

\begin{document}

\title{Higher-twist mechanism and inclusive gluon production in pion-proton collisions}
\author{A.~I.~Ahmadov$^{1,2}$~\footnote{E-mail: ahmadovazar@yahoo.com}}
\author{C.~Aydin$^{3}$~\footnote{E-mail: coskun@ktu.edu.tr}}
\author{R.~Myrzakulov$^{4}$~\footnote{E-mail: $\mbox{rmyrzakulov}\mbox{@gmail.com}$}}
\author{O.~Uzun$^{3}$~\footnote{E-mail: $\mbox{oguzhan}_-\mbox{deu@hotmail.com}$}}
\affiliation {$^{1}$ Fachbereich C, Bergische Universit\"at
Wuppertal, 42097 Wuppertal, Germany} \affiliation {$^{2}$ Department
of Theoretical Physics, Baku State University, Z. Khalilov st. 23,
AZ-1148, Baku, Azerbaijan} \affiliation {$^{3}$ Department of
Physics, Karadeniz Technical University, 61080, Trabzon, Turkey}
\affiliation {$^{4}$ Eurasion International Center for Theoretical
Physics and Department of General Theoretical Physics, Eurasian
National University, Astana 010008, Kazakhstan}

\date{\today}

\begin{abstract}
   We calculate the contribution of the higher-twist Feynman diagrams
to the large-$p_T$  inclusive gluon production cross-section in $\pi
p$ collisions in case of the running coupling and frozen coupling
approaches within perturbative and holographic QCD. The structure of
infrared renormalon singularities of the higher-twist subprocess
cross-section is obtained and  the resummed higher-twist cross-
sections (Borel sum) with the ones obtained in the framework of the
frozen coupling approach and leading-twist cross-section are
compared and analyzed.
\end{abstract}

\pacs{12.38.-t, 13.60.Le, 14.40.Aq, 13.87.Fh}

\keywords{Higher-twist, pion distribution amplitude, infrared
renormalons}

\maketitle

\section{\bf Introduction}

It is well known that Quantum Chromodynamics (QCD) is the
fundamental theory of the strong interactions.Therefore in order to
describe the structure and dynamical properties of hadrons at the
amplitude level many researchers have been studying QCD.  The
hadronic distribution amplitude in terms of internal structure
degrees of freedoms plays a crucial role in QCD process predictions.

One of the basic problems in QCD is choosing the renormalization
scale in running coupling constant $\alpha_{s}(Q^2)$. In principle,
in perturbative QCD (pQCD) calculations, the argument of the running
coupling constant in both the renormalization and factorization
scale $Q^2$ should be taken as equal to the square of the momentum
transfer of a hard gluon in a corresponding Feynman diagram
~\cite{Brodsky1}. In the pQCD, the physical information of the
inclusive gluon production is obtained efficiently; therefore, it
can be directly compared to the experimental data.

It should be noted that problem the existence of the higher-twist
contribution is not yet settled. Also it is necessary to study the
difference in the leading-twist results for the frozen and running
coupling constant approaches and compare it with that of the
higher-twist.

The aim of this study is calculation and analysis of the inclusive
gluon production in the pion-proton collisions using the frozen and
running coupling constant approaches. Using this approaches, the
higher twist effects have been already calculated by many
authors~\cite{Bagger,Bagger1,Baier,Sadykhov,Ahmadov1,Ahmadov2,
Ahmadov3,Ahmadov4,Ahmadov5,Ahmadov6,Ahmadov7,Ahmadov8,Ahmadov9,Ahmadov10,Gupta}.

The calculation and  analysis of the higher-twist effects on the
dependence of the pion distribution amplitude in inclusive gluon
production at $\pi p$ collision within holhographic and pQCD
approaches are   the interesting  research problems. In our
previously study~\cite{Ahmadov10}, using the principle of maximum
conformality (PMC) and Brodsky-Lepage-Mackenzie (BLM) approaches, we
have calculated the contribution of the higher-twist mechanism to
the large-$p_T$ inclusive gluon production cross section in $\pi p$
collisions ~\cite{Ahmadov10}. In this study, we compute the
contribution of the higher-twist effects to an inclusive gluon
production cross-section by using various pion distribution
amplitudes from holographic and pQCD, and applying the frozen and
running coupling~\cite{Agaev} constant approaches in order to
compute the effects of the infrared renormalons. We also estimate
and perform comparisons of the leading and the higher-twist
contributions. Studying various properties of the hadron infrared
renormalon effects in detail is one of the essential and actual
problems in the
pQCD~\cite{Hooft,Mueller,Mueller1,Zakharov,Beneke,Greiner,Contopanagos,
Kataev1,Kataev2}. This approach was also employed
before~\cite{Ahmadov3,Ahmadov4,Ahmadov5,Ahmadov6,Ahmadov7,Ahmadov8,Ahmadov9}
to calculate the inclusive meson production in $pp$ and
$\gamma\gamma$ collisions.

  The contents of the paper is as follows.  The  related formulas for the
calculation of the contributions of the higher-twist  and
leading-twist diagrams are provided in the next section. The
formulas and analysis of the higher twist effects on the dependence
of the pion distribution amplitudes  by the running  coupling
constant approach are presented in Section~\ref{ir}, and the
numerical results for the cross-section and discussion of the
dependence of the cross-section on the pion distribution amplitudes
are presented in Section~\ref{results}. Finally, our conclusions and
the highlights of the study are listed in Section~\ref{conc}.

\section{HIGHER-TWIST AND LEADING-TWIST CONTRIBUTIONS TO INCLUSIVE GLUON PRODUCTION}
\label{ht}

The higher-twist Feynman diagrams for the inclusive gluon production
in the pion-proton collision $\pi p \to g X$ are shown in Fig.1. For
the process $\pi p \to g X$, we write invariant amplitude as in the
form (as called by Brodsky-Lepage formula~\cite{Lepage2} )
\begin{equation}
M(\hat s,\hat
t)=\int_{0}^{1}{dx_1}\int_{0}^{1}dx_2\delta(1-x_1-x_2)\Phi_{M}(x_1,x_2,Q^2)T_{H}(x_1,x_2;Q^2,\mu_{R}^2,\mu_{F}^2)
\end{equation}
where $T_H$ is  the sum of the graphs contributing to the
hard-scattering part of the subprocess. For the  higher-twist, the
subprocess is taken as $\pi q_{p} \to g q$, which contributes to
$\pi p \to g X$,  where $q_{p}$ is a constituent of the initial
proton target. As seen from Fig.1, the processes  $\pi^{+} p \to g
X$ and $\pi^{-} p \to g X$ arise from subprocesses as $\pi^{+} d_{p}
\to g u$ and $\pi^{-} u_{p} \to g d$, respectively.

The production of the hadronic gluon  in the large transverse
momentum  is available at the high energy, especially at the Large
Hadron Collider. Finally, hadronic gluon is a product of the
hard-scattering processes, before hadronization. In the final state,
this hadronic gluon  is fragmented to hadron. The main dynamical
properties of the gluon, which carried one part of the four
momentum, are close to the parent parton. In order to understand the
parton kinematics one should consider the gluon production
process~\cite{Owens}.
  The higher-twist cross-section for $\pi p \to g X$ process has the form:
\begin{equation}
E\frac{d\sigma}{d^{3}p}(\pi p \to g X )=\int_{0}^{1}dx \delta(\hat
s+\hat t+\hat u)\hat s G_{q/{p}}(x, Q^2)\frac
{1}{\pi}\frac{d\sigma}{d\hat t}(\pi q_{p} \to g q),
\end{equation}
where $G_{q/p}(x, Q^2)$ is the quark distribution function  inside
a proton.

For higher-twist subprocess $\pi q_{p} \to g q$, the Mandelstam
invariant variables  are written in the form:
\begin{equation}
\hat s=(p_1+p_{g})^2=(p_2+p_{\pi})^2,\quad \hat
t=(p_{g}-p_{\pi})^2,\quad \hat u=(p_1-p_{\pi})^2.
\end{equation}
Then the parton-level cross-section within running coupling constant
method becomes
\begin{equation}
\frac{d\sigma}{d\hat t}(\pi q_p \rightarrow g q)=\frac {256\pi^2}
{81{\hat s}^2}\left[D(\hat s,\hat u)\right]^2\left(-\frac{\hat
t}{{\hat s}^2}-\frac{\hat t}{{\hat u}^2}\right)\, ,
\end{equation}
where
\begin{equation}
D(\hat s,\hat u)=\int_{0}^{1}dx
\alpha_{s}^{3/2}(Q_1^2)\left[\frac{\Phi_{\pi}(x,Q_{1}^2)}{x(1-x)}\right]+\int_{0}^{1}dx
\alpha_{s}^{3/2}(Q_2^2)\left[\frac{\Phi_{\pi}(x,Q_{2}^2)}{x(1-x)}\right].
\end{equation}

By the way, it must be denoted that as a special case  we can
directly get the result ~\cite{Berger} from Eq(2.4) by the applying
the frozen coupling constant approximation.

If we use the BLM approach for Fig.1 the transfer momentum of the
hard gluon for $s$ and  $u$ channels get the form
\begin{equation}
Q_{1}^2=(1-x)\hat s \,\,\,\,\,\,{\rm and} \,\,\,\,\,\,Q_{2}^2=-x\hat u,
\end{equation}
respectively.

In the soft regions $x\rightarrow 0$ and $x\rightarrow 1$  (for $u$
and $s$ channels), integrals (2.5) diverge, therefore in these
regions for their calculations some regularization methods of
$\alpha_{s}(Q^2)$ are needed.  One of the simple method is called
frozen coupling constant approximation for the regularization these
singularity.

Although the frozen QCD coupling constant was introduced a long time
ago, it is also interesting and important  in these
days~\cite{Curci,Curci1,Greco1,Greco2,Webber1,Webber2,Badelek,Ciafaloni,
Kotikov}. The origin of it comes from the divergent infrared
behavior of the well-known renormalization group expression for
$\alpha_{s}$. For this reason, it is used as a constant in the
infrared domain. Another reason for introducing the frozen coupling
is the pQCD coupling. Also, the effects of running $\alpha_{s}$
should be taken into account in all calculations. However, it makes
some QCD calculations very difficult. If we want to estimate it
approximately, it can be very convenient to use some effective
coupling which minimizes the running of $\alpha_{s}$ in the
perturbative region. For getting an agreement with experimental
data, the values of the frozen coupling are usually fixed from
purely phenomenological considerations. The frozen coupling is
frequently used in combination  with other phenomenological
parameters to describe hadronic processes. We can come across fixed
$\alpha_{s}$ very often  in various calculations done in the
framework of the leading logarithmic approximation where most
important logarithmic contributions are totally resummed while
$\alpha_{s}$ considered as fixed parameter and its argument  is set
off a posteriori from physical considerations. The solution of the
Schwinger-Dyson equations can be also another method for
investigating the infrared behavior of the gluon and ghost
propagators and for the running coupling constant at low
energies~\cite{Roberts}. Although early studies of the
Schwinger-Dyson equations  for the gluon propagator are very
singular in the infrared~\cite{Mandelstam, Brown,Brown1}, other
studies found infrared finite propagators. For example, in Ref.
~\cite{Cornwall} the gluon acquires a dynamical mass $m_{g}^2$, and
the other is extensively discussed in Refs.~\cite{Alkofer, Smekal}
when the gluon propagator goes to zero when the momentum
$Q^2\rightarrow 0$. In both cases, the freezing of coupling constant
appears in the infrared. In the case where squared momentum of a
hard gluon gets the form $Q^2 \to Q^2+m_{g}^2$, for running coupling
constant leads to $\alpha_{s}(Q^2)\to \alpha_{s}(Q^2+m_{g}^2)$. Here
$m_{g}^2$ is interpreted as an effective dynamical gluon mass or
fictitious mass of gluon.  By frozen coupling constant approach for
squared  of transfer momentum of the hard gluon in single meson
photoproduction, $\gamma p\rightarrow MX$ is taken as $Q_{1}^2=\hat
s/2$ and $Q_{2}^2= -\hat u/2$~\cite{Bagger}. Additionally, we also
come across other  physical properties of frozen coupling constant
in the confinement mechanism suggested in
Refs.~\cite{Parisi,Schwinger} as $(1+1)$ dimensional Quantum
Electrodyanamics .

There are few forms of the pion distribution amplitude available in
the literature. In the present  numerical calculations, we use
several choices, such as the asymptotic distribution amplitude
derived in pQCD evalution~\cite{Lepage1}, the
Vega-Schmidt-Branz-Gutsche-Lyubovitskij (VSBGL) distribution
amplitude~\cite{ Vega}, distribution amplitudes predicted by
 AdS/CFT~\cite{Brodsky2,Brodsky3}, the Chernyak-Zhitnitsky(CZ)~\cite{chernyak}, the
Bakulev-Mikhailov-Stefanis (BMS)~\cite{Bakulev,Bakulev1} and pion
distribution amplitudes in which Gegenbauer coefficients $C_2$ and
$C_4$ are extracted from BELLE experiment ~\cite{Braun,Uehara}:
\begin{equation}
\Phi_{asy}(x,Q^2\rightarrow\infty)=\sqrt{3}f_{\pi}x(1-x),
\label{asy}
\end{equation}
\begin{equation}
\Phi_{VSBGL}^{hol}(x,\mu_{0}^2)=\frac{A_1k_1}{2\pi}\sqrt{x(1-x)}exp\left(-\frac{m^2}{2k_{1}^2x(1-x)}\right),
\end{equation}
\begin{equation}
\Phi^{hol}(x,\mu_{0}^2)=\frac{4}{\sqrt{3}\pi}f_{\pi}\sqrt{x(1-x)},
\end{equation}
\begin{equation}
\Phi_{CZ}(x,\mu_{0}^2)=\Phi_{asy}(x)\left[C_{0}^{3/2}(2x-1)+\frac{2}{3}C_{2}^{3/2}(2x-1)\right],
\end{equation}
\begin{equation}
\Phi_{BMS}(x,\mu_{0}^2)=\Phi_{asy}(x)\left[C_{0}^{3/2}(2x-1)+0.20C_{2}^{3/2}(2x-1)-0.14C_{4}^{3/2}(2x-1)\right],
\end{equation}
\begin{equation}
\Phi_{BELLE}(x,\mu_{0}^2)=\Phi_{asy}(x)\left[C_{0}^{3/2}(2x-1)+0.12C_{2}^{3/2}(2x-1)+0.08C_{4}^{3/2}(2x-1)\right],
\label{BELLE}
\end{equation}
here $C_{n}^{\lambda}(2x-1)$ are Gegenbauer polynomials.

Substituting  Eq.(2.4) into Eq.(2.2), then the differential  cross-
section for the process $\pi p \to g X$ takes the form
~\cite{Berger}
\begin{equation}
E\frac{d\sigma}{d^{3}p}(\pi p\to g X )=\frac{s}{s+u} xG_{q/p}(x,
Q^2)\frac {256\pi} {81{\hat s}^2}\left[D(\hat s,\hat
u)\right]^2\left(-\frac{\hat t}{{\hat s}^2}-\frac{\hat t}{{\hat
u}^2}\right).
\end{equation}
It should be noted that, as seen from Eqs.(2.4) and (2.13), the
higher-twist cross-section is linear with respect to $\hat t$, so
the cross-section vanishes, if the scattering angle between the
final gluon and incident pion is approximately equal to zero. From
Eq.(2.13), we see that the higher-twist cross-section proportional
to $\hat s^{-3}$, which is equivalent to the higher-twist
contributions to the $\pi p\to gX$ cross-section have the form of
$p_{T}^{-6}f(x_{F},x_{T})$.

In the expression Eq.(2.6), we fixed the variable $x$ by taking it
as mean value. So, average values for $x$ we take $\overline x =
1/2$. Thus, for the calculations higher-twist cross-sections within
frozen coupling constant approach we substitute $\overline Q^2=\hat
s/2$ and $\overline Q^2=-\hat u/2$ in Eq.(2.13) for the transfer
momentum of the hard gluon, respectively.

 The extracting of higher-twist contribution from the inclusive
gluon production cross-section is also complicated. One can also
consider the comparison of higher-twist corrections with
leading-twist contributions. For the leading-twist subprocess in the
inclusive gluon production, we take $q\bar{q} \to g\gamma$ as a
subprocess of the quark-antiquark annihilation. The differential
cross-section for subprocess $q\bar{q} \to g\gamma$ is
\begin{equation}
\frac{d\sigma}{d\hat t}(q\bar{q} \to g\gamma)=\frac{8}{9}\pi\alpha_E
\frac{e_{q}^2}{{\hat s}^2}\left(\alpha_s(-\hat u)\frac{\hat t}{\hat
u}+\alpha_s(-\hat t)\frac{\hat u}{\hat t}\right).
\end{equation}

As is seen from Eq.(2.14), leading-twist cross-section strongly
depends of the running coupling constant where the running coupling
constant depends on the transfer momentum. However, running coupling
constant depends on the channels of the process. Here running
coupling have been evaluated in the momentum subtraction scheme, for
momentum scales $u$ and $t$, which  define the off-shell momenta
carried by the quark propagators.

The leading-twist  cross-section for production of inclusive gluon
is~\cite{Berger1}
\begin{equation}
\Sigma_{M}^{LT}\equiv E\frac{d\sigma}{d^{3}p}(\pi p \to g X
)=\int_{0}^{1}dx_{1} \int_{0}^{1}dx_{2} \delta(\hat s+\hat t+\hat u)
G_{\overline{q}/{M}}(x_{1},Q_{1}^2)G_{q/{p}}(x_{2},Q_{2}^2)\frac
{\hat s}{\pi}\frac{d\sigma}{d\hat t}(q\bar{q} \to g\gamma),
\end{equation}
where
$$
 \hat{s}=x_{1}x_{2}s,\ \hat{t}=x_{1}t,\  \hat{u}=x_{2}u.
$$

Finally, leading-twist contribution to the large-$p_{T}$ gluon
production cross-section in the process $\pi p\to g X$ is
\begin{equation}
\Sigma_{M}^{LT}\equiv E\frac{d\sigma}{d^{3}p}(\pi p \to g X
)=\int_{0}^{1}dx_{1} \frac{1}{x_1s+u}
G_{\overline{q}/{M}}(x_{1},Q_{1}^2)G_{q/{p}}(1-x_{1},Q_{2}^2)\frac
{\hat s}{\pi}\frac{d\sigma}{d\hat t}(q\bar{q} \to g\gamma).
\end{equation}

\section{HIGHER TWIST MECHANISM  WITHIN PERTURBATIVE AND HOLOGRAPHIC QCD
AND THE ROLE INFRARED RENORMALONS}
\label{ir}

The main object of this study is the calculations of  the higher-
twist cross-section with running coupling constant approach within
holographic and pQCD and renormalon effect's contribution to the
cross-section, and also comparisons between higher-twist cross-
sections which are calculated by the running coupling constant
method and the principle maximum conformality approach. It should be
noted that, in the exclusive processes, the coupling constant
$\alpha_{s}$ runs not only due to the loop integration but also the
integration in the process amplitude over the light-cone momentum
fraction of hadron constituents. Therefore, it is worth noting that
the renormalization scale according to Fig.1 should be chosen equal
to $\mu_{R_1}^2=Q_{1}^2=(1-x)\hat s$, and
$\mu_{R_2}^2=Q_{2}^2=-x\hat u$. The integral in Eq.(2.5) in the
framework of the running coupling approach takes the form
\begin{equation}
D(\mu_{R}^2)=\int_{0}^{1}\frac{\alpha^{3/2}_{s}((1-x)\hat
s)\Phi_{M}(x,\mu_{F}^2)dx}{x(1-x)}+\int_{0}^{1}\frac{\alpha^{3/2}_{s}(-x\hat
u)\Phi_{M}(x,\mu_{F}^2)dx}{x(1-x)} .
\end{equation}

At the leading order of pQCD calculations the hard scattering
amplitude $T_{H}(x_1,x_2;Q^2, \mu_{R}^2,\mu_{F}^2)$ does not depend
on the factorization scale $\mu_{F}^2$, but strongly depends on
$\mu_{R}^2$. The one-loop QCD correction to the hard scattering
amplitude $T_{H}(x_1,x_2;Q^2, \mu_{R}^2,\mu_{F}^2)$ generates its
explicit dependence on the scales $\mu_{F}^2$ and $\mu_{R}^2$  .
Also, it should be noted that the scales $\mu_{F}^2$ and $\mu_{R}^2$
are independent of each other so they can be chosen autonomously.

As we noted above in the regions $x\rightarrow 0$ and $x\rightarrow
1$ the integral $(3.1)$ diverges, because in this region the running
coupling constants  $\alpha_{s}((1-x)\hat s)$ and $\alpha_{s}(-x\hat
u)$ have the infrared singularity. In other words, the singularity
of the integrand  of x=0 and x=1 is due to only by
$\alpha_{s}((1-x)\hat s)$ and $\alpha_{s}(-x\hat u)$. Thus, for the
regularization of the integral, by expressing the running coupling
at scaling variable $\alpha_{s}( \mu_{R}^2)$, we use renormalization
group equation with the  fixed $\alpha_{s}(\hat s)$ and
$\alpha_{s}(-\hat u)$ for  $s$ and $u$ channels, respectively. The
solution of renormalization group equation for the running coupling
$\alpha\equiv\alpha_{s}/\pi$ is in the form ~\cite{Contopanagos}

\begin{equation}
\frac{\alpha(\lambda)}{\alpha}=\left[1+\alpha
\frac{\beta_{0}}{4}\ln{\lambda}\right]^{-1}.
\end{equation}
Then, for $\alpha_{s}((1-x)\hat s)$, we get
\begin{equation}
\alpha((1-x)s)=\frac{\alpha_{s}}{1+ln(1-x)/t}
\end{equation}
where $t=4\pi/\alpha_{s}(Q^2)\beta_{0}=4/\alpha\beta_{0}$.

If we insert  Eq.(3.3) into Eq.(3.1), we obtain
$$
D(\hat s,\hat u)=\int_{0}^{1}dx\frac{\alpha^{3/2}_{s}((1-x)\hat
s)\Phi_{\pi}(x,Q_{1}^2)}{x(1-x)}+
\int_{0}^{1}dx\frac{\alpha^{3/2}_{s}(-x\hat
u)\Phi_{\pi}(x,Q_{2}^2)}{x(1-x)}
$$
\begin{equation}
=\alpha^{3/2}_{s}(\hat s)t^{3/2}_{1}\int_{0}^{1}dx
\frac{\Phi_{\pi}(x,Q_{1}^2)}{x(1-x)(t_{1}+\ln\lambda_1)^{3/2}} +
\alpha^{3/2}_{s}(-\hat u)t^{3/2}_{2}\int_{0}^{1}dx
\frac{\Phi_{\pi}(x,Q_{2}^2)}{x(1-x)(t_{2}+\ln\lambda_2)^{3/2}}
\end{equation}
where $t_1=4\pi/\alpha_{s}(\hat s)\beta_{0}$ and
$t_2=4\pi/\alpha_{s}(-\hat u)\beta_{0}$.

Although the integral (3.4)  still has  singularity, this expression
can be transformed to a more convenient form by the change of
variable as, $z=\ln\lambda$. Then the singularity in (3.4)
disappears and we obtain

\begin{equation}
D(\hat s,\hat u)= \alpha^{3/2}_{s}(\hat s) t^{3/2}_{1} \int_{0}^{1}
\frac{\Phi_{\pi}(x,Q_{1}^2)dx}{x(1-x)(t_{1}+z_{1})^{3/2}}+
 \alpha^{3/2}_{s}(-\hat u) t^{3/2}_{2} \int_{0}^{1}
\frac{\Phi_{\pi}(x,Q_{2}^2)dx}{x(1-x)(t_{2}+z_{2})^{3/2}}.
\end{equation}
After applying the integral representation of $1/(t+z)^{\nu}$
~\cite{Zinn-Justin,Erdelyi},
\begin{equation}
\frac{1}{(t+z)^{\nu}}=\frac{1}{\Gamma(\nu)}\int_{0}^{\infty}e^{-(t+z)u}
u^{\nu-1}du,  Re\nu>0
\end{equation}
Equation (3.5) becomes
\begin{eqnarray}
D(\hat s,\hat u)=\frac{\alpha^{3/2}_{s}(-\hat s)
t_1^{3/2}}{\Gamma(\frac{3}{2})} \int_{0}^{1} \int_{0}^{\infty}
\frac{\Phi_{\pi}(x,Q_{1}^2)e^{-(t_{1}+z_{1})u}u^{1/2}du dx}{x(1-x)}+ \nonumber \\
+\frac{\alpha^{3/2}_{s}(-\hat u) t^{3/2}_2}{\Gamma(\frac{3}{2})}
\int_{0}^{1} \int_{0}^{\infty}
\frac{\Phi_{\pi}(x,Q_{2}^2)e^{-(t_{2}+z_{2})u}u^{1/2}du dx}{x(1-x)},
\end{eqnarray}
Then the Eq.(3.7) can be written, for $\Phi^{hol}(x,Q^2)$, as
%
\ba
D(\hat s,\hat u)=\frac{32\sqrt{\pi} f_{\pi}
}{\beta_{0}\sqrt{3\beta_{0}}\Gamma{(\frac{3}{2})}} \int_{0}^{\infty}
du e^{-t_{1}u}u^{1/2}B\left(\frac{1}{2},\frac{1}{2}-u\right)+ \nn \\
+\frac{32\sqrt{\pi}f_{\pi} }{\beta_{0}\sqrt{3\beta_{0}}\Gamma{(\frac{3}{2})}}
\int_{0}^{\infty} du
e^{-t_{2}u}u^{1/2}B\left(\frac{1}{2},\frac{1}{2}-u\right),
\ea
for $\Phi_{asy}(x,Q^2\rightarrow\infty)$ distribution amplitude, as
%
\begin{equation}
D(\hat s,\hat u)=\frac{8\pi\sqrt{3\pi}
f_{\pi}}{\beta_{0}\sqrt{\beta_{0}}\Gamma{(\frac{3}{2})}}
\int_{0}^{\infty}du e^{-t_{1}u} \left[\frac{u^{1/2}}{1-u}\right]
+\frac{8\pi\sqrt{3\pi}
f_{\pi}}{\beta_{0}\sqrt{\beta_{0}}\Gamma{(\frac{3}{2})}}
\int_{0}^{\infty}du e^{-t_{2}u} \left[\frac{u^{1/2}}{1-u}\right],
\end{equation}
for $\Phi_{CZ}(x,Q^2)$ distribution amplitude, as
\ba
D(\hat s,\hat u)&=&\frac{8\pi\sqrt{3\pi}
f_{\pi}}{\beta_{0}\sqrt{\beta_{0}}\Gamma{(\frac{3}{2})}}
\int_{0}^{\infty}du e^{-t_{1}u}u^{1/2}
\left[\frac{1}{1-u}+0.84\left[\frac{4}{1-u}-\frac{20}{2-u}+\frac{20}{3-u}\right]\left(\frac{\alpha_{s}(Q_{1}^2)}{\alpha_{s}(\mu_{0}^2)}
\right)^{\frac{50}{81}}\right]+
\nn \\
&&\frac{8\pi\sqrt{3\pi}f_{\pi}}{\beta_{0}\sqrt{\beta_{0}}\Gamma{(\frac{3}{2})}}
\int_{0}^{\infty}du e^{-t_{2}u}u^{1/2}
\left[\frac{1}{1-u}+0.84\left[\frac{4}{1-u}-\frac{20}{2-u}+\frac{20}{3-u}\right]\left(\frac{\alpha_{s}(Q_{2}^2)}{\alpha_{s}(\mu_{0}^2)}
\right)^{\frac{50}{81}}\right]. \nn \\
\ea
and for $\Phi_{BMS}(x,Q^2)$ distribution amplitude, as
\ba
D(\hat s,\hat u)&=&\frac{8\pi\sqrt{3\pi}
f_{\pi}}{\beta_{0}\sqrt{\beta_{0}}\Gamma{(\frac{3}{2})}}
\int_{0}^{\infty}du e^{-t_{1}u}u^{1/2}
\biggl[\frac{1}{1-u}+0.30\biggl[\frac{4}{1-u}-\frac{20}{2-u}+\frac{20}{3-u}\biggr]\biggl(\frac{\alpha_{s}(Q_{1}^2)}{\alpha_{s}(\mu_{0}^2)}
\biggr)^{\frac{50}{81}}-
\nn \\
&&0.2625\biggl[\frac{8}{1-u}-
\frac{120}{2-u}+\frac{560}{3-u}-\frac{1112}{4-u}+\frac{1008}{5-u}-\frac{336}{6-u}\biggr]\biggl(\frac{\alpha_{s}(Q_{1}^2)}{\alpha_{s}(\mu_{0}^2)}
\biggr)^{364/405}\biggr]+ \nn \\
&&\frac{8\pi\sqrt{3\pi}
f_{\pi}}{\beta_{0}\sqrt{\beta_{0}}\Gamma{(\frac{3}{2})}}
\int_{0}^{\infty}du e^{-t_{2}u}u^{1/2}
\biggl[\frac{1}{1-u}+0.30\biggl[\frac{4}{1-u}-\frac{20}{2-u}+\frac{20}{3-u}\biggr]\biggl(\frac{\alpha_{s}(Q_{2}^2)}{\alpha_{s}(\mu_{0}^2)}
\biggr)^{\frac{50}{81}}-
\nn \\
&&0.2625\biggl[\frac{8}{1-u}-
\frac{120}{2-u}+\frac{560}{3-u}-\frac{1112}{4-u}+\frac{1008}{5-u}-\frac{336}{6-u}\biggr]\biggl(\frac{\alpha_{s}(Q_{2}^2)}{\alpha_{s}(\mu_{0}^2)}
\biggr)^{364/405}\biggr].
\ea
where $B(\alpha,\beta)$ is Beta function and $u$ is the Borel
parameter.

The structure of the infrared renormalon poles in Eqs.(3.8)-(3.11)
strongly depend on the distribution amplitudes of the pion. The
integrals in Eqs.(3.8)-(3.11) are divergent.Therefore there are
regularized by means of the principal value prescription by using
running coupling constant. In the subsequent calculation and figures
the higher-twist cross sections obtained using the running coupling
and frozen coupling constant approaches are denoted by
$(\Sigma_{g}^{HT})^{res}$ and $(\Sigma_{g}^{HT})^{0}$, respectively.

\section{NUMERICAL RESULTS AND DISCUSSION}
\label{results}

We discuss the numerical results for higher-twist and renormalon
mechanism with higher-twist contributions calculated in the context
of the running  and frozen coupling approaches on the dependence of
the chosen pion distributions amplitudes in the inclusive gluon
production process. For the numerical calculations, we take
supprocess $\pi^{+} d_{p} \to g u$ and $\pi^{-} u_{p} \to g d$ for
$\pi^{+} p \to g X$ and $\pi^{-} p \to g X$ process, respectively.

Inclusive direct gluon production represents a significant test case
in which higher-twist terms dominate those of leading-twist in
certain kinematic domains. For the dominant leading-twist subprocess
for the gluon production, we take the quark-antiquark annihilation
$q\bar{q} \to \gamma g$.  In the numerical calculations, for the
quark distribution functions inside the pion and proton we used
expressions as given in Refs. \cite{nam,watt}, respectively.

Results obtained in our calculations are visualized in Figs. 2-23.
In all figures we represent the choice of pion distribution
amplitudes Eqs.(\ref{asy})-(\ref{BELLE}) by different line types:
$\Phi_{asy}(x)$ as solid black line, $\Phi^{hol}(x)$ as dashed red
line, $\Phi_{VSBGL}^{hol}(x)$ as dotted blue line,
$\Phi_{CZ}(x,Q^2)$ as dash-dot magenta line, $\Phi_{BMS}(x,Q^2)$  as
dash-double dot olive line, and $\Phi_{BELLE}(x,Q^2)$ as short dash
navy line. Firstly, it is very interesting to compare the
higher-twist cross sections obtained within  holographic QCD with
ones obtained within the perturbative QCD and also with the
leading-twist cross-section. In Figs. 2 and 3, we show higher-twist
cross-sections $(\Sigma_{g}^{HT})^{0}, (\Sigma_{g}^{HT})^{res}$
calculated in the context of the frozen (frozen cross-section) and
running coupling constant (resummed cross-section) approaches as a
function of the gluon transverse momentum $p_{T}$ for the pion
distribution amplitudes presented in Eqs.(2.6)-(2.11) at $y=0$. It
is seen from Figs. 2 and 3 that the higher-twist cross-section is
monotonically decreasing with an increase in the transverse momentum
of the gluon. In the region $2\,\,GeV/c<p_T<30\,\,GeV/c$, the
resummed cross-sections of the process $\pi^{+}p \to g X$ decrease
in the range between $3.172\cdot10^{-6}\mu b/GeV^{2}$ to
$4.912\cdot10^{-16}\mu b/GeV^{2}$. In Figs. 4-7, we show
$(\Sigma_{g}^{HT})^{res}/(\Sigma_{g}^{HT})^{0}$,$(\Sigma_{g}^{HT})^{res}/(\Sigma_{g}^{HT})^{PMC}$,
$(\Sigma_{g}^{HT})^{0}/(\Sigma_{g}^{LT})$ and
$(\Sigma_{g}^{HT})^{res}/(\Sigma_{g}^{LT})$ for the process $\pi^{+}
p \to g X$  as a function of $p_{T}$ for the pion distribution
amplitudes presented in Eqs.(\ref{asy})-(\ref{BELLE}) at $y=0$. We
see in Fig. 4, that in the region $15\,\,GeV/c<p_T<22\,\,GeV/c$, the
ratio $(\Sigma_{g}^{HT})^{res}/(\Sigma_{g}^{HT})^{0}$  for
$\Phi_{CZ}(x,Q^2)$ is enhanced by about two orders of magnitude
relative to one for $\Phi_{asy}(x)$. However, the enhancement is one
order of magnitude for $\Phi^{hol}(x)$ and half an order for
$\Phi_{BMS}(x,Q^2)$ and $\Phi_{BELLE}(x,Q^2)$ pion distribution
amplitudes. For the  present the distinction between running
coupling constant and principle maximum conformality approaches in
Fig. 5 is shown where the
$(\Sigma_{g}^{HT})^{res}/(\Sigma_{g}^{HT})^{PMC}$ ratio of
higher-twist cross-sections is calculated by running coupling
constant method and principle maximum conformality approach as a
function of the transverse momentum of the gluon  $p_{T}$. For full
analysis in Table \ref{table1} we present numerically values of the
ratio $(\Sigma_{g}^{HT})^{res}/(\Sigma_{g}^{HT})^{PMC}$ as a
function of the transverse momentum of the gluon $p_{T}$. In Figs.6
and 7, we show the dependence of the ratios
$(\Sigma_{g}^{HT})^{0}$/$(\Sigma_{g}^{LT})$ and
$(\Sigma_{g}^{HT})^{res}$/$(\Sigma_{g}^{LT})$ on the gluon $p_{T}$
for the pion distribution amplitudes presented in Eqs.(2.6)-(2.11).
It is seen that in the region $10\,\,GeV/c<p_T<25\,\,GeV/c$
leading-twist cross-sections are enhanced by about four orders of
magnitude relative to the higher-twist cross-sections calculated in
the frozen coupling constant approach, but in some regions they are
enhanced by about three orders of magnitude relative to the resummed
higher-twist cross-section for $\Phi_{CZ}(x,Q^2)$. In Figs. 8-11, we
have depicted higher-twist cross-sections $(\Sigma_{g}^{HT})^{0}$,
$(\Sigma_{g}^{HT})^{res}$ and ratios
$(\Sigma_{g}^{HT})^{res}/(\Sigma_{g}^{HT})^{0}$,
$(\Sigma_{g}^{HT})^{0}/(\Sigma_{g}^{LT})$  and
$(\Sigma_{g}^{HT})^{res}/(\Sigma_{g}^{LT})$ as a function of $p_{T}$
at $\sqrt s=62.4\,\,GeV$ for the process $\pi^{-}p\to g X$. In Figs.
8 and 9, we display the dependence of higher-twist cross-sections
$(\Sigma_{g}^{HT})^{0}$, $(\Sigma_{g}^{HT})^{res}$ as a function of
$p_{T}$  for pion distribution amplitudes presented in
Eqs.(\ref{asy})-(\ref{BELLE}) at $y=0$. From the figures, we see
that the higher-twist cross-section is monotonically decreasing with
an increase of $p_{T}$. In this process, in the region
$2\,\,GeV/c<p_T<30\,\,GeV/c$, the resummed cross-sections decrease
in the range between $4.556\cdot10^{-6}\mu b/GeV^{2}$ and
$6.056\cdot10^{-14}\mu b/GeV^{2}$. In Figs. 10 and 11, we display
the dependence of the ratios
$(\Sigma_{g}^{HT})^{0}$/$(\Sigma_{g}^{LT})$ and
$(\Sigma_{g}^{HT})^{res}$/$(\Sigma_{g}^{LT})$ for the process
$\pi^{-} p \to g X$ as a function of  $p_{T}$ for the pion
distribution amplitudes presented in Eqs.(\ref{asy})-(\ref{BELLE})
at $y=0$. It should be noted, that ratio
$(\Sigma_{g}^{HT})^{res}$/$(\Sigma_{g}^{HT})^{0}$ for the $\pi^{+} p
\to g X$ process is identical to the ratio
$(\Sigma_{g}^{HT})^{res}$/$(\Sigma_{g}^{HT})^{0}$  for the process
$\pi^{-} p \to g X$. In Fig.10, we show the dependence of the ratio
$(\Sigma_{g}^{HT})^{res}$/$(\Sigma_{g}^{LT})$, as a function of
$p_{T}$ for the pion distribution amplitudes presented in
Eqs.(\ref{asy})-(\ref{BELLE}) at $y=0$. We also see from the figure
that in the region $10\,\,GeV/c<p_T<30\,\,GeV/c$, leading-twist
cross-sections are enhanced by about three orders of magnitude
relative to the frozen cross-sections for $\Phi^{hol}(x)$,
$\Phi_{asy}(x)$, $\Phi_{CZ}(x,Q^2)$, $\Phi_{BMS}(x,Q^2)$ and
$\Phi_{BELLE}(x,Q^2)$ at $y=0$. But the enhancement is about 4
orders of magnitude relative to the frozen cross section for
$\Phi_{VSBGL}^{hol}(x)$ distribution amplitude.

From Figs. 12-19, the dependences of higher-twist cross-sections
$(\Sigma_{g}^{HT})^{0}$, $(\Sigma_{g}^{HT})^{res}$, ratios
$(\Sigma_{g}^{HT})^{res}$/$(\Sigma_{g}^{HT})^{0}$,
$(\Sigma_{g}^{HT})^{res}$/$(\Sigma_{g}^{HT})^{PMC}$  and
$(\Sigma_{g}^{HT})^{res}$/$(\Sigma_{g}^{LT})$  are shown for the
processes $\pi^{+} p \to \gamma X$ and $\pi^{-} p \to \gamma X$ as a
function of the rapidity of the gluon $y$ at the transverse momentum
of the gluon $p_T=4.9\,\, GeV/c$. It is seen from figures in Figs.12
and 13, that frozen and resummed cross sections for all distribution
amplitudes  of pion have two maxima, where the first maximum is
approximately at the point $y=-2$ and second maximum is
approximately at the point $y=2$. But, the leading-twist cross-
section only has one maximum at $y=1.5$. Notice that the
distribution amplitude of frozen and resummed cross-sections for
$\Phi_{CZ}(x,Q^2)$ are enhanced by about half and two orders of
magnitude relative to all other distribution amplitudes. In
Figs.14-19, dependences of higher-twist cross-sections
$(\Sigma_{g}^{HT})^{0}$, $(\Sigma_{g}^{HT})^{res}$, ratios
$(\Sigma_{g}^{HT})^{res}$/$(\Sigma_{g}^{HT})^{0}$,
$(\Sigma_{g}^{HT})^{res}$/$(\Sigma_{g}^{HT})^{PMC}$ and
$(\Sigma_{g}^{HT})^{res}$/$(\Sigma_{g}^{LT})$ for processes $\pi^{+}
p \to g X$ and $\pi^{-} p \to g X$ are displayed as a function of
the rapidity of the gluon $y$ at $p_T=4.9\,\, GeV/c$. For full
analysis in Table \ref{table2}, we present numerically values of the
ratio $(\Sigma_{g}^{HT})^{res}/(\Sigma_{g}^{HT})^{PMC}$  as a
function of the rapidity of the gluon $y$ at $p_T=4.9\,\, GeV/c$.

Angular distributions in the higher-twist cross sections
$(\Sigma_{g}^{HT})^{0}$, $(\Sigma_{g}^{HT})^{res}$ for processes
$\pi^{\pm} p \to g X$  are presented in Figs. 20-23. As is seen from
the figures, cross-sections vary slowly and smoothly with the angle
of the scattering. However, angular distributions are very sensitive
to the choice of the pion distribution amplitude.

As is seen from Figs.5 and 15, and also from Tables I and II, the
magnitude of the higher-twist cross-section calculated by running
coupling constant approach in common is enhanced by about 0.5-1
order of magnitude relative to the PMC cross-sections (cross-section
is calculated in the  principle of maximum conformality approach).
Main reason for this  is depend from phenomenological factors. As
noted above in the exclusive  processes the coupling constant runs
not only due to loop integration, but also because of the
integration in the process amplitude over the longitudinal momentum
fractions of hadron constituents. Thus, the exclusive processes have
two independent sources of power corrections to their
characteristics; the loop integration and the integration over the
longitudinal momentum fractions of quarks and gluons. It is worth
noting that the latter source exists even at the leading order of
pQCD, when the amplitude of the exclusive process depends on $\alpha
_{s}$. PMC scheme as is fixed factorization scheme and  differ from
frozen scheme with factor $e^{-5/3}$. But, in the running coupling
constant approach all contributions from quark and gluon line to the
cross-section must be taken into account. Thus, we can conclude that
the main reason of the difference between  resummed cross-section
(cross-section is calculated in the running coupling constant
approach) and the relative PMC cross-section  is more dependent on
this factor. Since only leading-twist diagrams are commonly
considered in the usual studies of the hadron-hadron collision, in
our calculations, magnitude of the leading-twist cross-section
enhanced frozen, PMC and resummed cross-sections. But, the magnitude
is the resummed cross-section the relative PMC cross-section is more
near to the leading-twist cross-section. Therefore, resummed cross-
section is more reliable.

Analysis of our calculations shows that $(\Sigma_{g}^{HT})^{0}$ and
$(\Sigma_{g}^{HT})^{res}$, frozen and resummed higher-twist cross-
sections, and ratios are sensitive to pion distribution amplitudes
as predicted in the holographic and pQCD.

We think that this feature of infrared renormalons may help
theoretical interpretations of the future experimental data for the
direct inclusive gluon production cross-section in the pion-proton
collisions. Higher-twist cross-section obtained in our study should
be observable at hadron collider.

\section{CONCLUSIONS}
\label{conc}

In this study, the inclusive single gluon production is calculated
via higher twist mechanism within perturbative and  holographic QCD.
In the calculation of the cross-sections the running  and frozen
coupling constant approaches are employed and infrared renormalon
poles in the cross-section expression are revealed. Infrared
renormalon induced divergences are regularized by means of the
principal value prescription and the Borel sum for the higher twist
cross-section is found. It is observed that the resummed
higher-twist cross-section differs from that found using the frozen
coupling approximation, especially in some regions, considerably.

Concerning the study of the higher-twist contribution, it is
primarily important to analyze its relative magnitude of
contribution compared to the leading-twist contribution, since only
leading-twist diagrams are commonly considered in usual studies of
the hadron-hadron collision. However, in our studies the difference
of the higher-twist results for the frozen and running coupling
constant approaches have been studied with importance. The following
results can be concluded from the experiments: the higher-twist
contributions to single gluon production cross-section in the
pion-proton collisions have important phenomenological consequences.
Therefore, they will be helpful for detailed investigation of the
dynamical properties of the nucleon. Also, the higher-twist gluon
production cross-section in the pion-proton collisions depends on
the form of the pion distribution amplitudes and may be used for
future study. Moreover, the contributions of renormalon effects
within holograpich QCD in these process are essential and may help
to analyze experimental results. We compared frozen and resumed
cross-sections of the direct gluon production in the  processes
$\pi^{-} p \to g X$ and $\pi^{+} p \to g X$. Our calculations show
in both cases running and frozen coupling constant approaches where
the inclusive gluon production cross-section for the process
$\pi^{-} p \to g X$ is suppress over the direct gluon production
cross-section of the process  $\pi^{+} p \to g X$. Note that the
direct gluon production spectrum can be measured with large
precision, so results obtained in this study will be helpful further
tests of the hadron dynamics at large $p_{T}$. As is seen from
Eqs.(2.4) and (3.7) higher-twist cross-sections in both cases are
proportional to the third power of $\alpha_{s}(Q^2)$, but the
leading-twist is linearly proportional to $\alpha_{s}(Q^2)$.
Therefore, their ratios strongly depend on the $\alpha_{s}^2(Q^2)$.

The cross-section which is calculated by running coupling constant
for all distribution amplitudes is enhanced by about 1-2 orders of
magnitude of cross-section which is calculated by the frozen
coupling constant method. As is seen from  Figs. 6, 7, 10 and 11,
the curves go up when $p_T$ is greater than 25 GeV. According to our
calculations, the main reason for it is the  approaching  to the
very nearing of singularity point $x\rightarrow  1$ of the running
coupling constant in this interval. Also, distinction is shown
between running coupling constant and principle maximum conformality
approaches.

Further investigations are needed in order to clarify the role of
higher-twist effects  in QCD. In hadron-hadron collisions, real
gluons at high transverse momentum can serve as a short distance
probe of the incident hadrons. Especially, the future experimental
measurements will provide further tests of the dynamics of
large-$p_T$ hadron production beyond the leading twist.

\section*{Acknowledgments}

One of the authors, A. I. Ahmadov is grateful to Prof. Robert
Harlander and  other members of the Department of Theoretical
Physics University of Wuppertal for provoding hospitality extended
to him in Wuppertal, where this work was carried out and to
Deutscher Akademischer Austausch Dienst (DAAD) for financial
support. The authors are indebted to V. M. Braun for discussions
about the properties of the pion distributions amplitudes. A. I.
Ahmadov is grateful for the financial support Baku State University
Grant No."50+50" (2014-2015). The work of C.Aydin is supported by
KTU under the research project no. BAP 2013/11641. We are also
grateful to A.~Mustafayev for carefully reading the paper and
providing useful comments.

\newpage
\begin{table}[h]
\begin{center}
\begin{tabular}{|c|c|c|c|c|c|} \hline
$p_{T},GeV/c$
&$\frac{(\Sigma_{g}^{HT})_{asy}^{res}}{(\Sigma_{g}^{HT})_{asy}^{PMC}}$
&
$\frac{(\Sigma_{g}^{HT})_{hol}^{res}}{(\Sigma_{g}^{HT})_{hol}^{PMC}}$
&
$\frac{(\Sigma_{g}^{HT})_{CZ}^{res}}{(\Sigma_{g}^{HT})_{CZ}^{PMC}}$&
$\frac{(\Sigma_{g}^{HT})_{BMS}^{res}}{(\Sigma_{g}^{HT})_{BMS}^{PMC}}$&$\frac{(\Sigma_{g}^{HT})_{BELLE}^{res}}{(\Sigma_{g}^{HT})_{BELLE}^{PMC}}$ \\
\hline
  2&0.63645&0.41676 &0,5223&4,5586&1,20161 \\ \hline
  4 &0.80708&1.11238&1,49909 &2,47698& 3,72963 \\ \hline
  6&0.83428&1.69304 &2,58183 & 2,58954& 6,73942 \\ \hline
  8&0.8428&2.19939&4,08479&2,95917&10,0357 \\ \hline
  10&0.84622&2.64643&6,37155&3,60209&13,1482  \\ \hline
  12&0.84818&3.0365&10,0184&4,64649&15,545 \\ \hline
  14&0.85958&3.3647&15,924&6,30331&16,9558 \\ \hline
  16&0.85075&3.62063&25,1232&8,80723&17,5175 \\ \hline
  18&0.8518&3.78841&37,3196&12,0757&17,6427 \\ \hline
  20&0.8528&3.84589&47,2563&14,8965&17,795 \\ \hline
  22&0.85377&3.76309&45,3689&14,9752&18,3262 \\ \hline
  24&0.85471&3.49973&31,618&11,9484&19,2869 \\ \hline
  26&0.85565&3.00164& 17,2571&8,38829&19,8528 \\ \hline
  28&0.85658&2.19545&8,3118&6,32575&17,4478 \\ \hline
  30&0.85752&0.98074&3,84996&7,25441&10,4326 \\ \hline
\end{tabular}
\end{center}
\caption{Numerically value of the  ratio
$(\Sigma_{g}^{HT})^{res}/(\Sigma_{g}^{PMC})$ in  the  process
$\pi^{+} p\to g X$ as a function of the transverse momentum of the
gluon $p_T$ at the c.m. energy $\sqrt s=62.4\,\, GeV$.}
\label{table1}
\end{table}

\begin{table}[h]
\begin{center}
\begin{tabular}{|c|c|c|c|c|c|} \hline
$y$
&$\frac{(\Sigma_{g}^{HT})_{asy}^{res}}{(\Sigma_{g}^{HT})_{asy}^{PMC}}$
&
$\frac{(\Sigma_{g}^{HT})_{hol}^{res}}{(\Sigma_{g}^{HT})_{hol}^{PMC}}$
&
$\frac{(\Sigma_{g}^{HT})_{CZ}^{res}}{(\Sigma_{g}^{HT})_{CZ}^{PMC}}$&
$\frac{(\Sigma_{g}^{HT})_{BMS}^{res}}{(\Sigma_{g}^{HT})_{BMS}^{PMC}}$&$\frac{(\Sigma_{g}^{HT})_{BELLE}^{res}}{(\Sigma_{g}^{HT})_{BELLE}^{PMC}}$\\
\hline
  -2.52&0,70335&2,75164&7,25081& 0,6715 & 11,1351 \\ \hline
  -2.22&0,1691&0,09216&0,03577 &0,11873& 0,09152 \\ \hline
  -1.98&0,19788&0,08752 &0,04549 &0,12802&0,11118 \\ \hline
  -1.62&0,31968&0,14862&0,11516& 2,05618 &0,29162 \\ \hline
  -1.32&0,47399&0,26103&0,26544&14,0557  &0,70125 \\ \hline
  -1.02&0,61431&0,426&0,50365& 2,85089 & 1,3795 \\ \hline
  -0.72&0,71688&0,64321&0,82097&2,18539 &  2,31447\\ \hline
  -0.42&0,78075&0,91273&1,21556& 2,18544 & 3,50047\\ \hline
  -0.12&0,81581&1,23771 &1,71207&2,37834 & 4,98013\\ \hline
  0.18&0,83309 &1,62501&2,38349& 2,68376 & 6,84887\\ \hline
  0.48&0,84078 &2,08299&3,40504&3,13548 & 9,20182 \\ \hline
  0.78&0,84381 &2,61508&5,218&3,90202 &11,9325 \\ \hline
  1.08&0,84481&3,20622&9,08747&5,47397 & 14,3304 \\ \hline
  1.38&0,84501 &3,79647&19,0914&9,30134& 15,1705\\ \hline
  1.68&0,84512 &4,23116&43,7879&18,109 &14,504 \\ \hline
  1.98&0.84586&4,16763&43,1949&19,44 & 14,7382\\ \hline
  2.28&0,84847&2,90664&9,06416& 7,92783&15,0257 \\ \hline
  2.52&0,9112&1,9231&8,9012&6,9834& 16,2354\\ \hline

\end{tabular}
\end{center}
\caption{Numerically values of the  ratio higher-twist cross
sections $(\Sigma_{g}^{HT})^{res}/(\Sigma_{g}^{HT})^{PMC}$
calculated by running coupling constant method and principle maximum
conformality approach as a function of the  rapidity  of the gluon
$y$, at the transverse momentum of the gluon  $p_T=4.9\,\, GeV/c$ at
the c.m. energy $\sqrt s=62.4\,\, GeV$.} \label{table2}
\end{table}

\newpage

\begin{figure}[!hbt]
\vskip -0.5cm \epsfxsize 15cm \centerline{\epsfbox{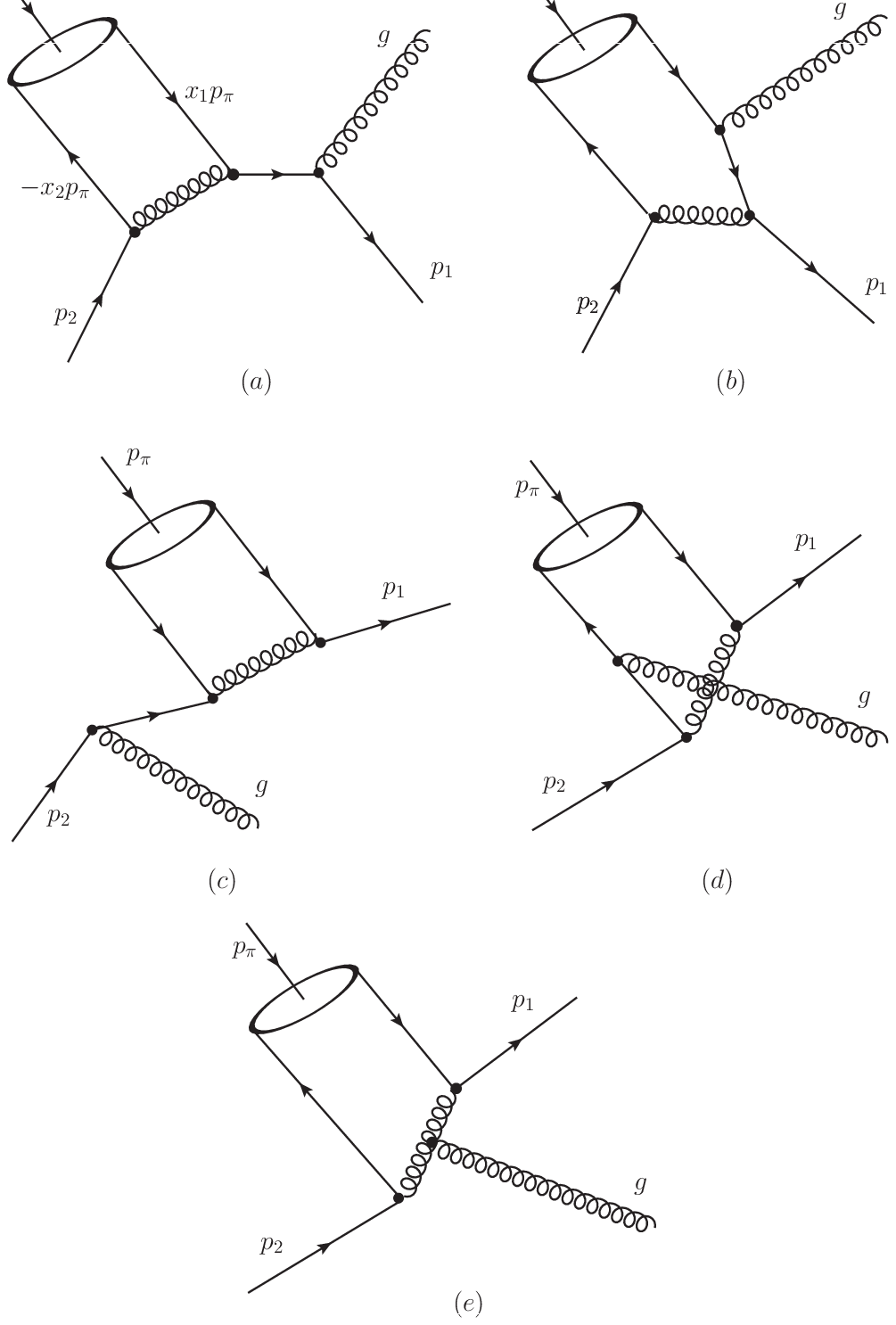}} \vskip
-2cm \caption{ Full set of QCD Feynman diagrams for higher-twist
subprocess $\pi q\to g q$.}
\label{Fig1}
\end{figure}

\begin{figure}[!hbt]
\vskip 1.2cm\epsfxsize 11.8cm \centerline{\epsfbox{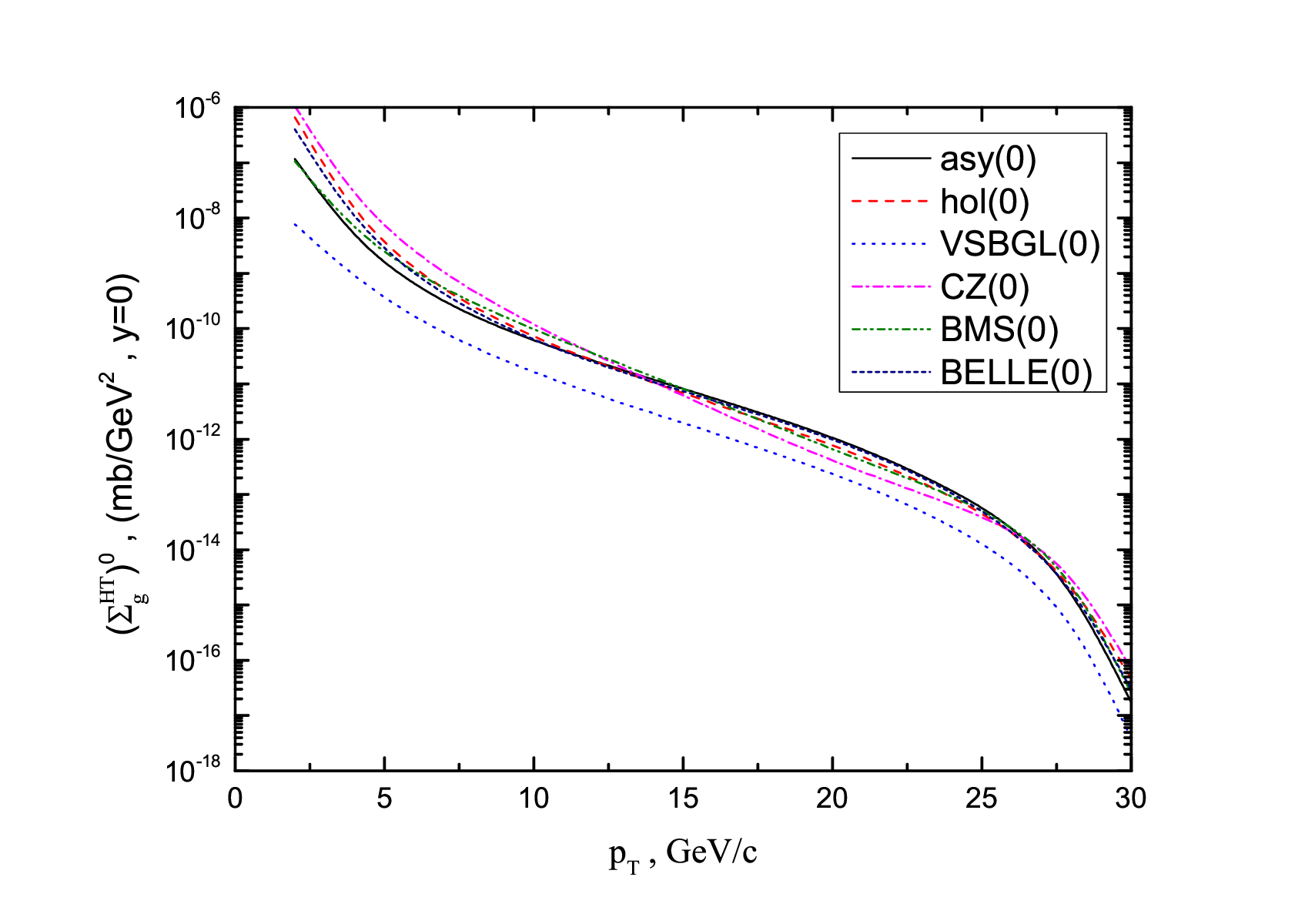}}
\vskip-0.2cm \caption{Higher-twist $\pi^{+} p\to g X$ inclusive
gluon production cross-section $(\Sigma_{g}^{HT})^{0}$ as a function
of the transverse momentum of the gluon $p_{T}$ at the c.m. energy
$\sqrt s=62.4\,\, GeV$.} \label{Fig2}
\end{figure}

\begin{figure}[!hbt]
\vskip 1.2cm \epsfxsize 11.8cm \centerline{\epsfbox{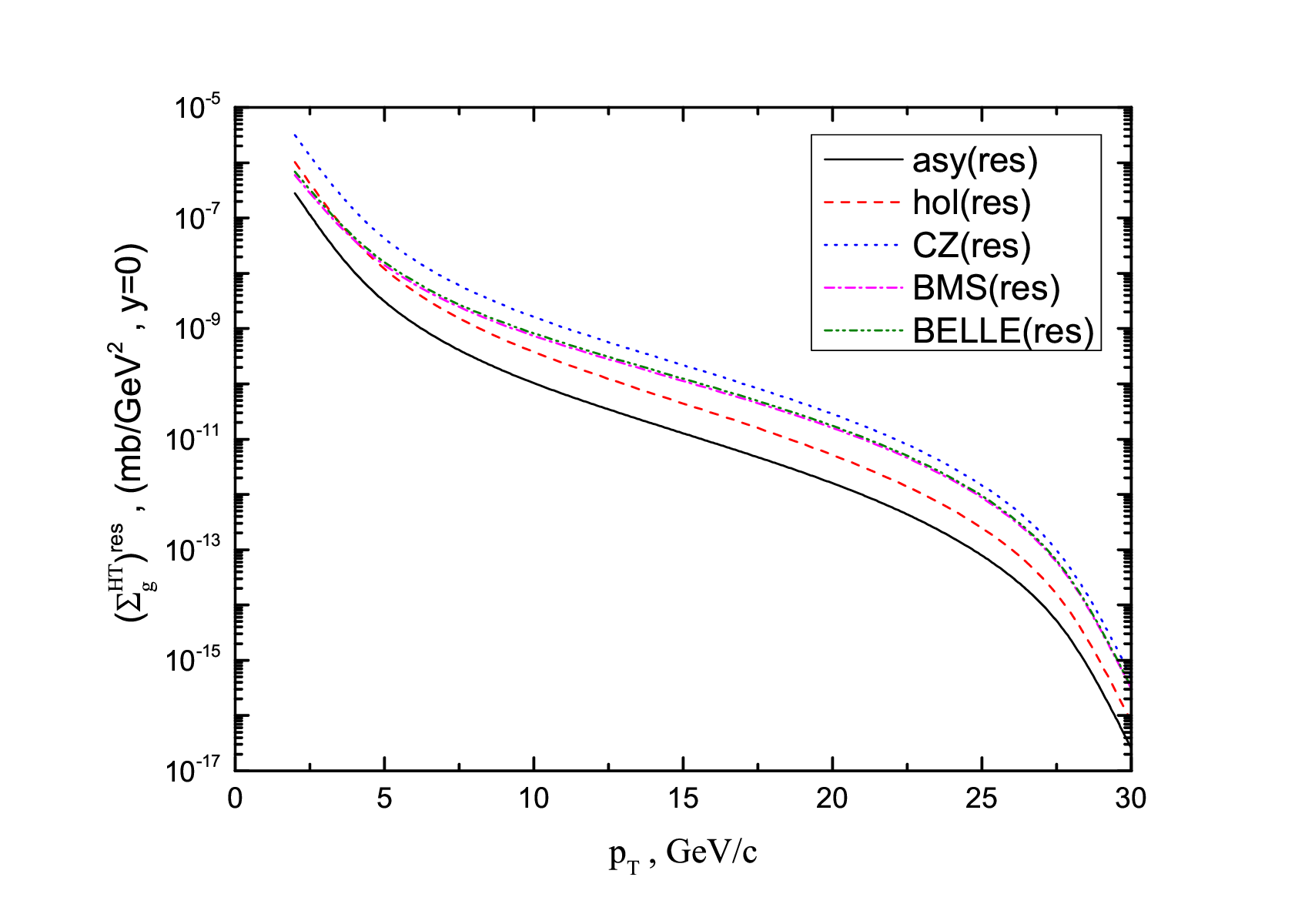}}
\vskip-0.2cm \caption{Higher-twist $\pi^{+} p\to g X$ inclusive
gluon production cross-section $(\Sigma_{g}^{HT})^{res}$ as a
function of the transverse momentum of the gluon $p_{T}$ at the
c.m.energy $\sqrt s=62.4\,\, GeV$.} \label{Fig3}
\end{figure}

\begin{figure}[!hbt]
\vskip -1.2cm\epsfxsize 11.8cm \centerline{\epsfbox{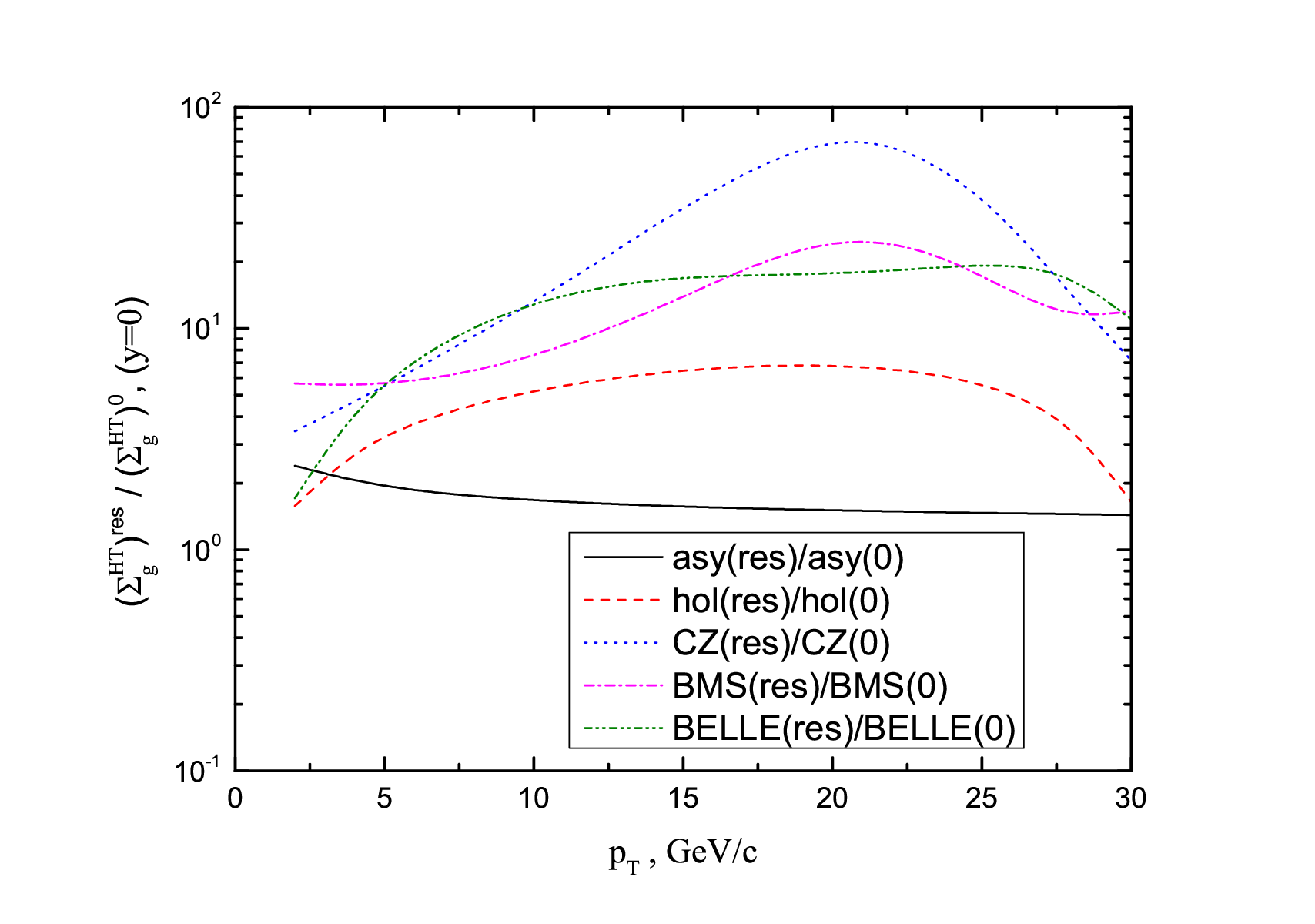}}
\vskip-0.2cm \caption{Ratios
$(\Sigma_{g}^{HT})^{res}/(\Sigma_{g}^{HT})^{0}$, in the process
$\pi^{+} p\to g X$, where higher-twist contribution are calculated
for the gluon rapidity $y=0$ at the c.m.energy $\sqrt s=62.4\,\,
GeV$ as function of the gluon transverse momentum, $p_{T}$.}
\label{Fig4}
\end{figure}

\begin{figure}[!hbt]
\vskip 1.2cm\epsfxsize 11.8cm \centerline{\epsfbox{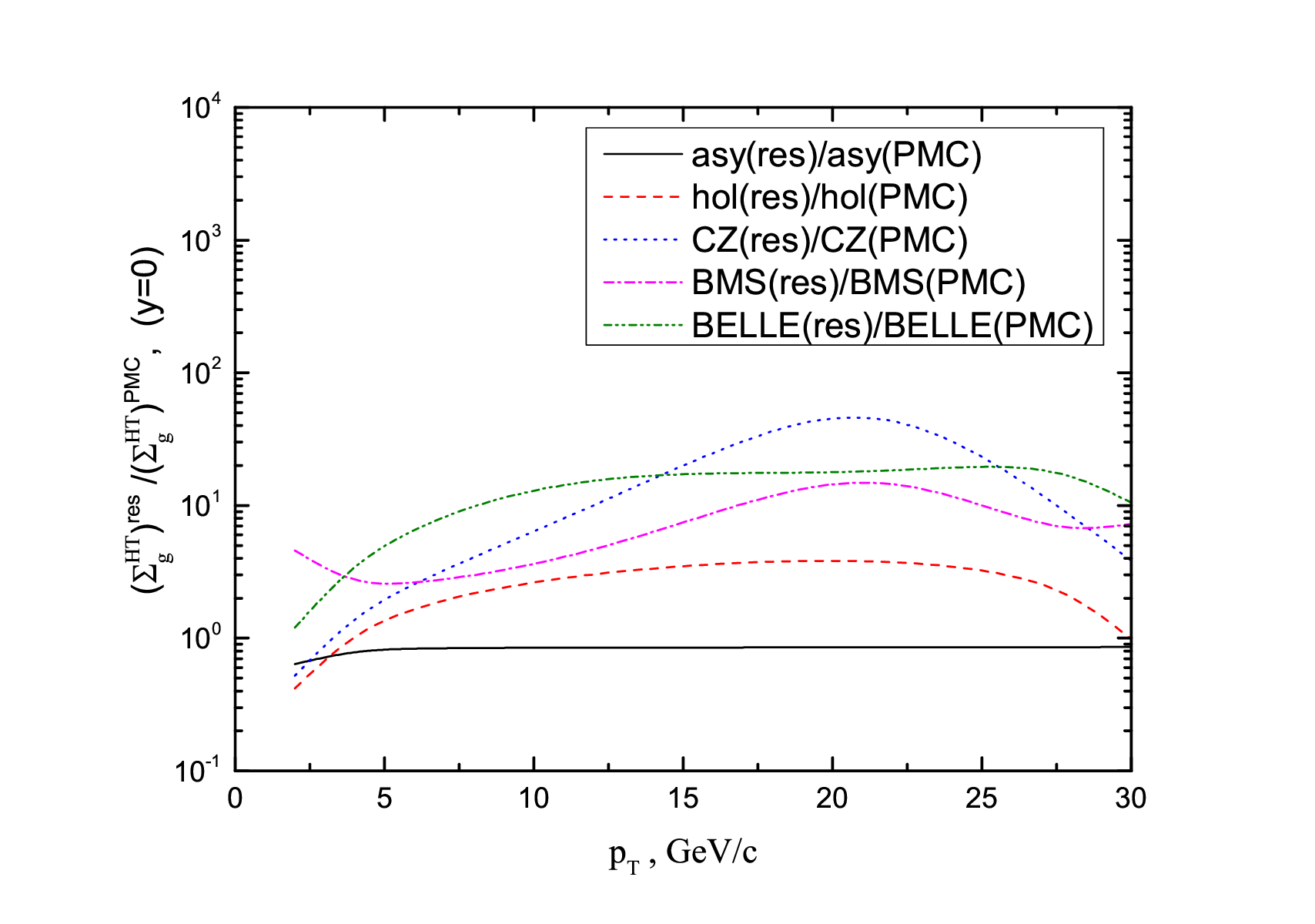}}
\vskip-0.2cm \caption{Ratios
$(\Sigma_{g}^{HT})^{res}/(\Sigma_{g}^{PMC})$ for the process
$\pi^{+} p\to g X$ as function of the transverse momentum of the
gluon at the c.m. energy $\sqrt s=62.4\,\, GeV$.} \label{Fig5}
\end{figure}

\begin{figure}[!hbt]
\vskip 1.2cm\epsfxsize 11.8cm \centerline{\epsfbox{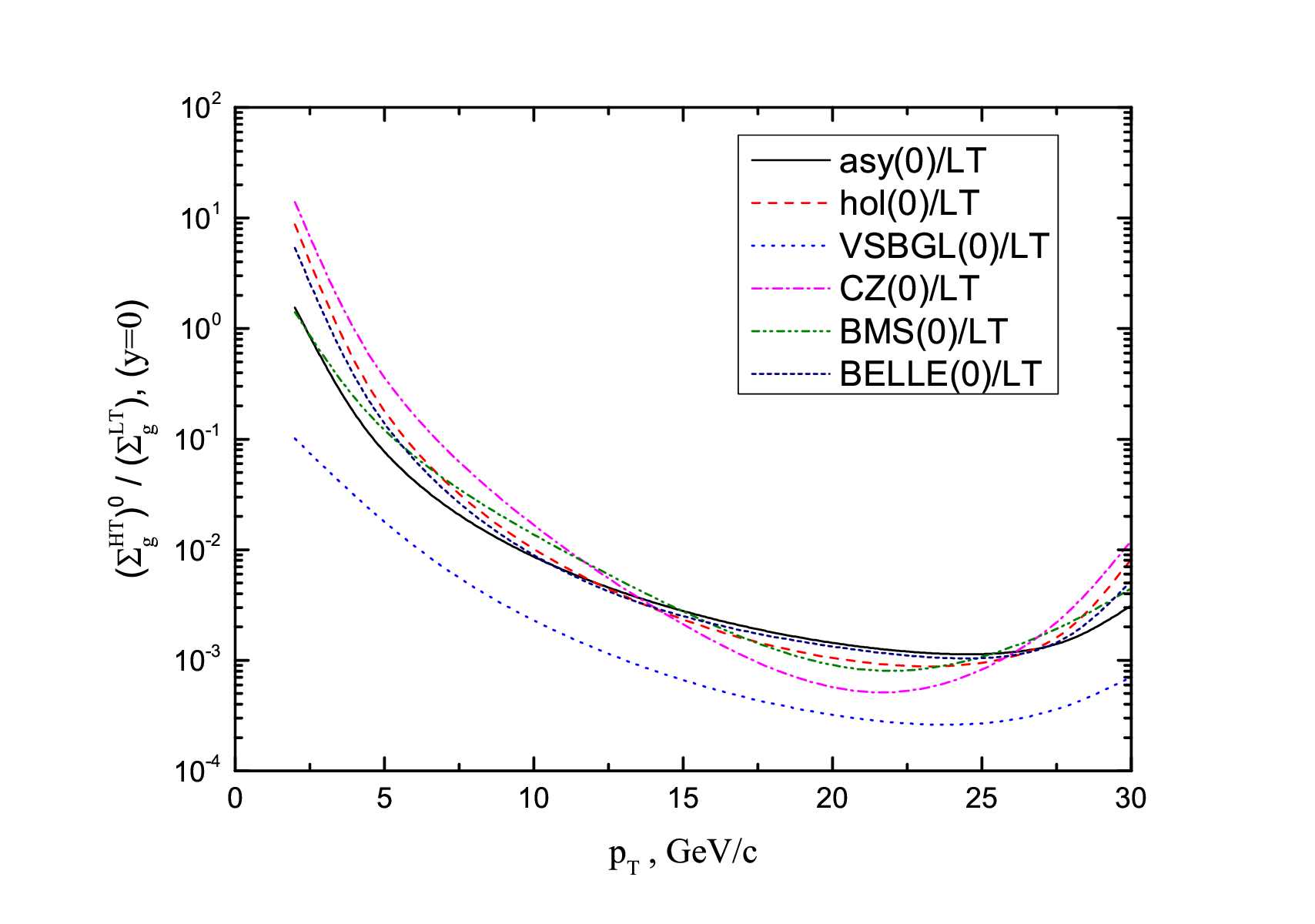}}
\vskip-0.2cm \caption{Ratios
$(\Sigma_{g}^{HT})^{0}/(\Sigma_{g}^{LT})$ for the process $\pi^{+}
p\to g X$ as function of the transverse momentum of the gluon at the
c.m. energy $\sqrt s=62.4\,\, GeV$.} \label{Fig6}
\end{figure}

\begin{figure}[!hbt]
\vskip -1.2cm\epsfxsize 11.8cm \centerline{\epsfbox{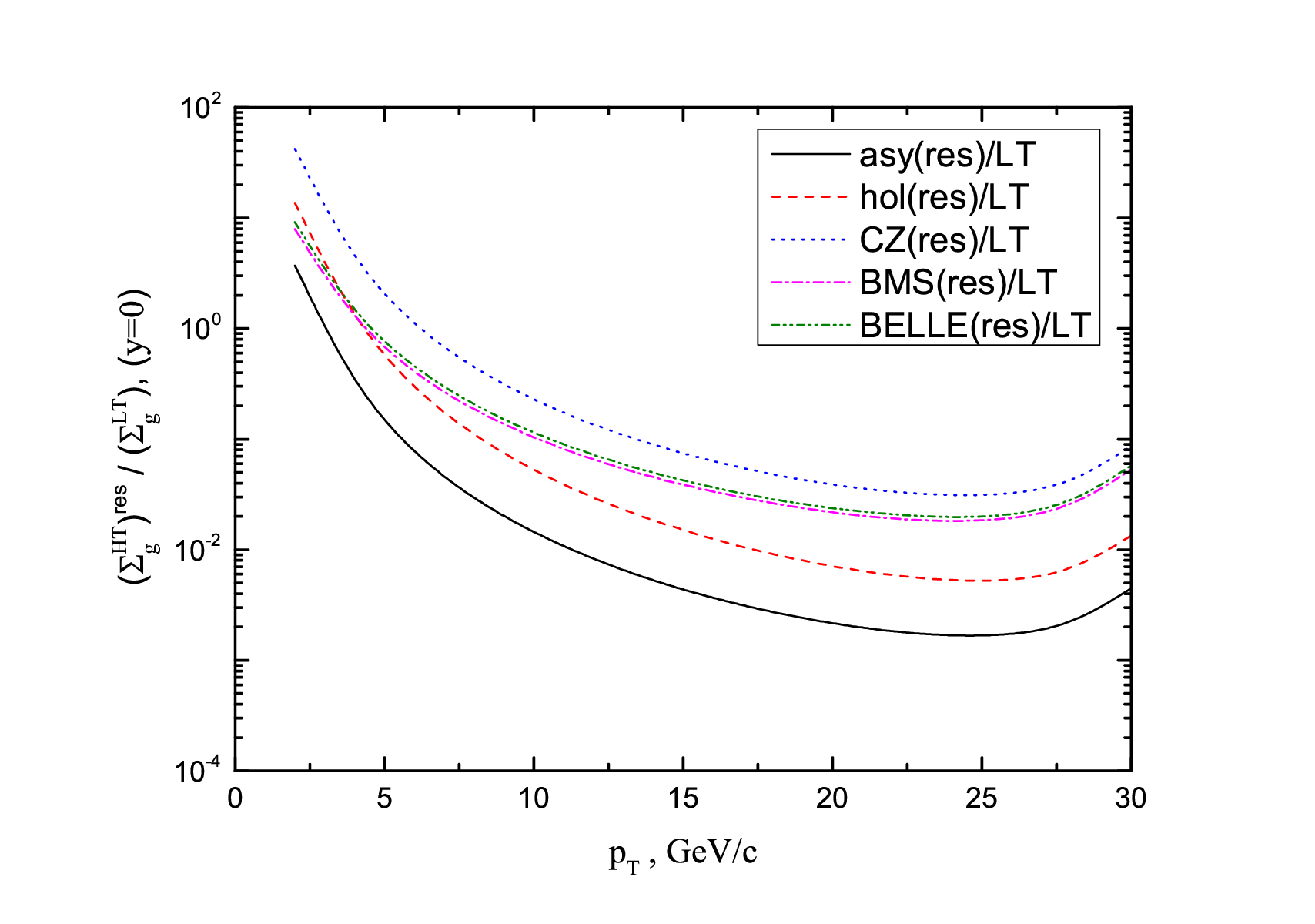}}
\vskip-0.2cm \caption{Ratios
$(\Sigma_{g}^{HT})^{res}/(\Sigma_{g}^{LT})$, in the process $\pi^{+}
p\to g X$, as a function of the transverse momentum of the gluon
$p_{T}$ at the c.m. energy $\sqrt s=62.4\,\, GeV$.} \label{Fig7}
\end{figure}

\begin{figure}[!hbt]
\vskip 1.2cm\epsfxsize 11.8cm \centerline{\epsfbox{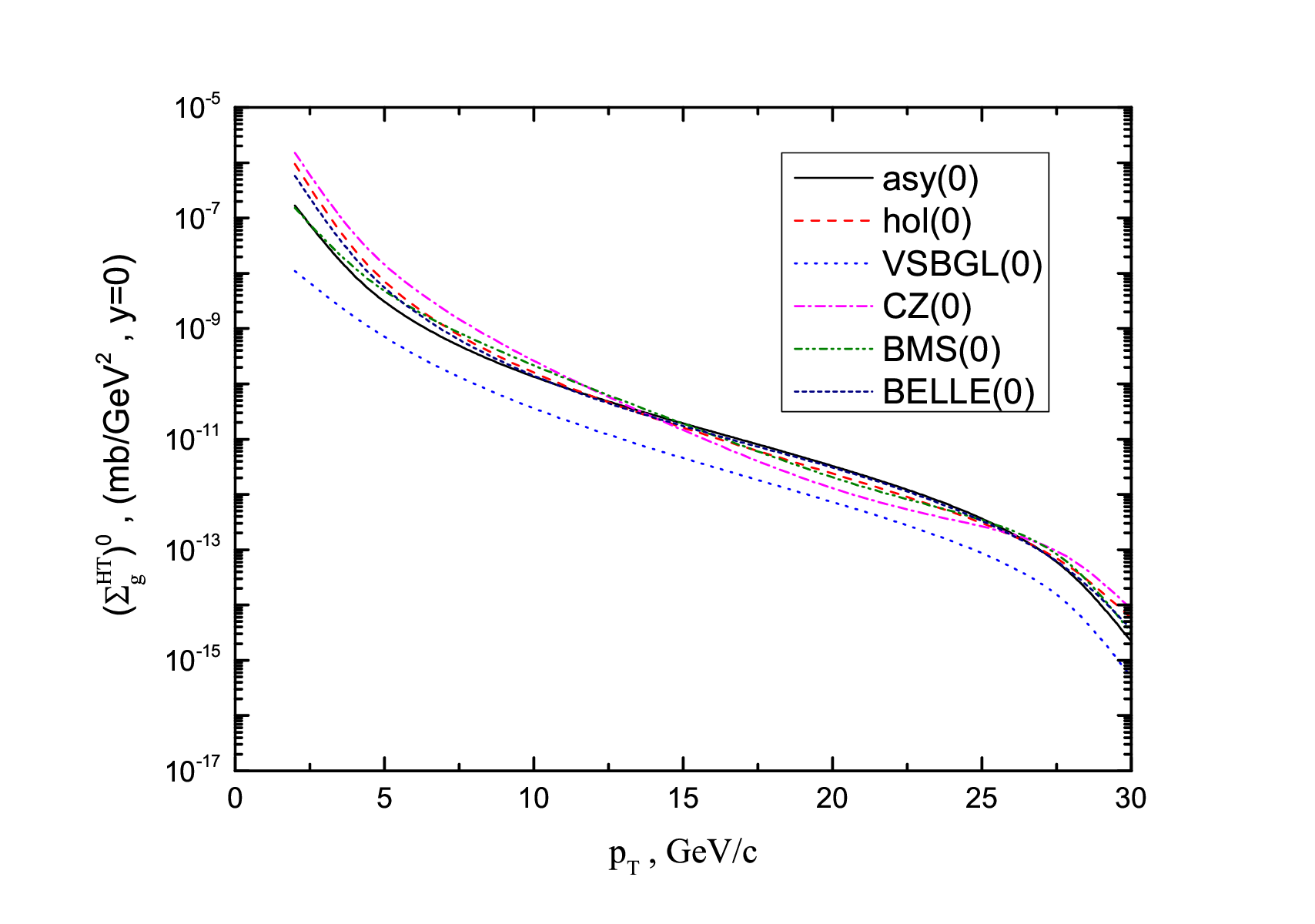}}
\vskip-0.2cm \caption{Higher-twist $\pi^{-} p\to g X$ inclusive
gluon production cross-section $(\Sigma_{g}^{HT})^{0}$ as function
of the transverse momentum of the gluon $p_{T}$  at the c.m. energy
$\sqrt s=62.4\,\, GeV$.} \label{Fig8}
\end{figure}

\begin{figure}[!hbt]
\vskip-1.2cm \epsfxsize 11.8cm \centerline{\epsfbox{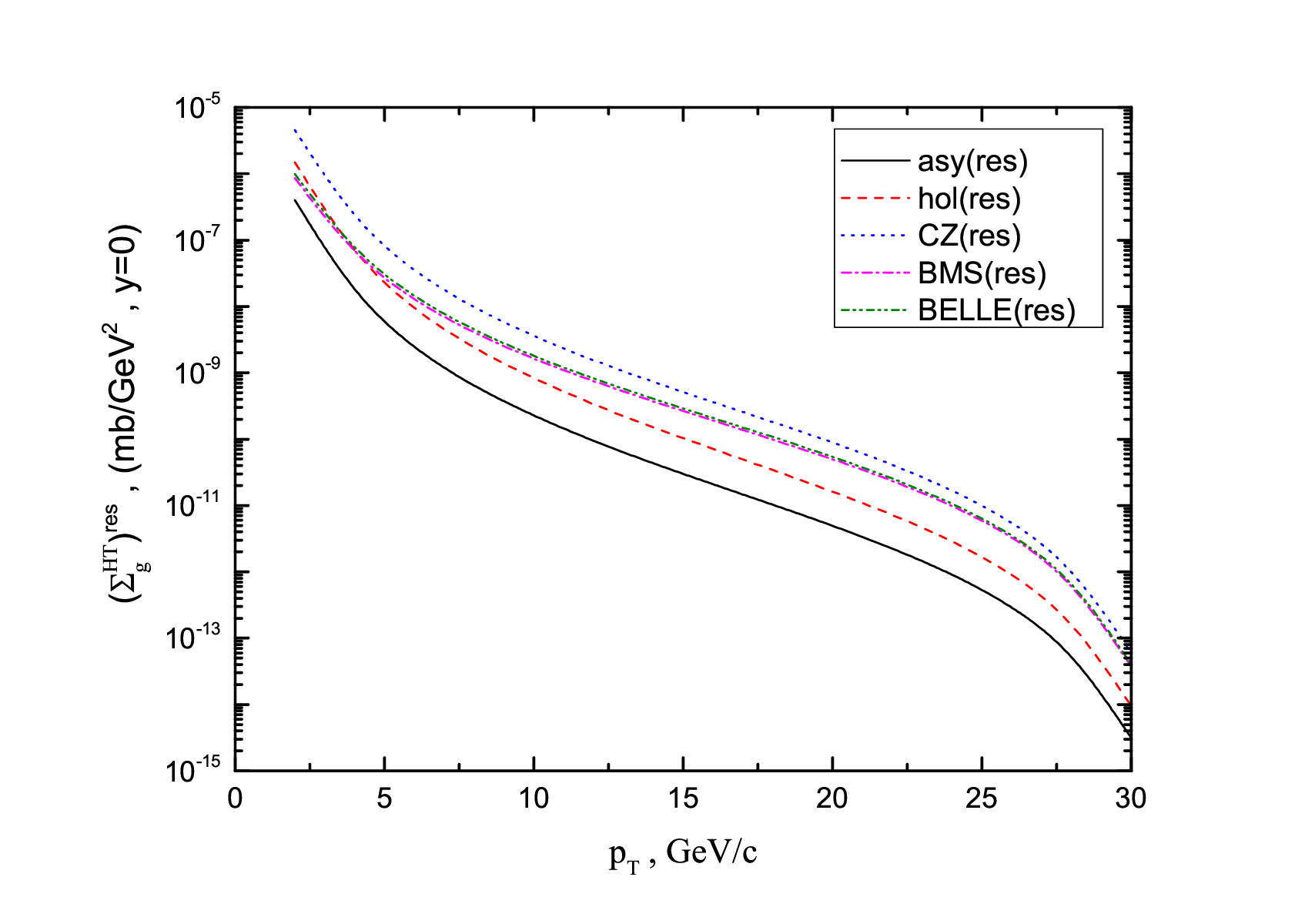}}
\vskip-0.2cm \caption{Higher-twist $\pi^{-} p\to g X$ inclusive
gluon production cross-section $(\Sigma_{g}^{HT})^{res}$ as function
of the transverse momentum of the gluon  $p_{T}$ at the c.m.energy
$\sqrt s=62.4\,\, GeV$.} \label{Fig9}
\end{figure}

\begin{figure}[!hbt]
\vskip-1.2cm \epsfxsize 11.8cm \centerline{\epsfbox{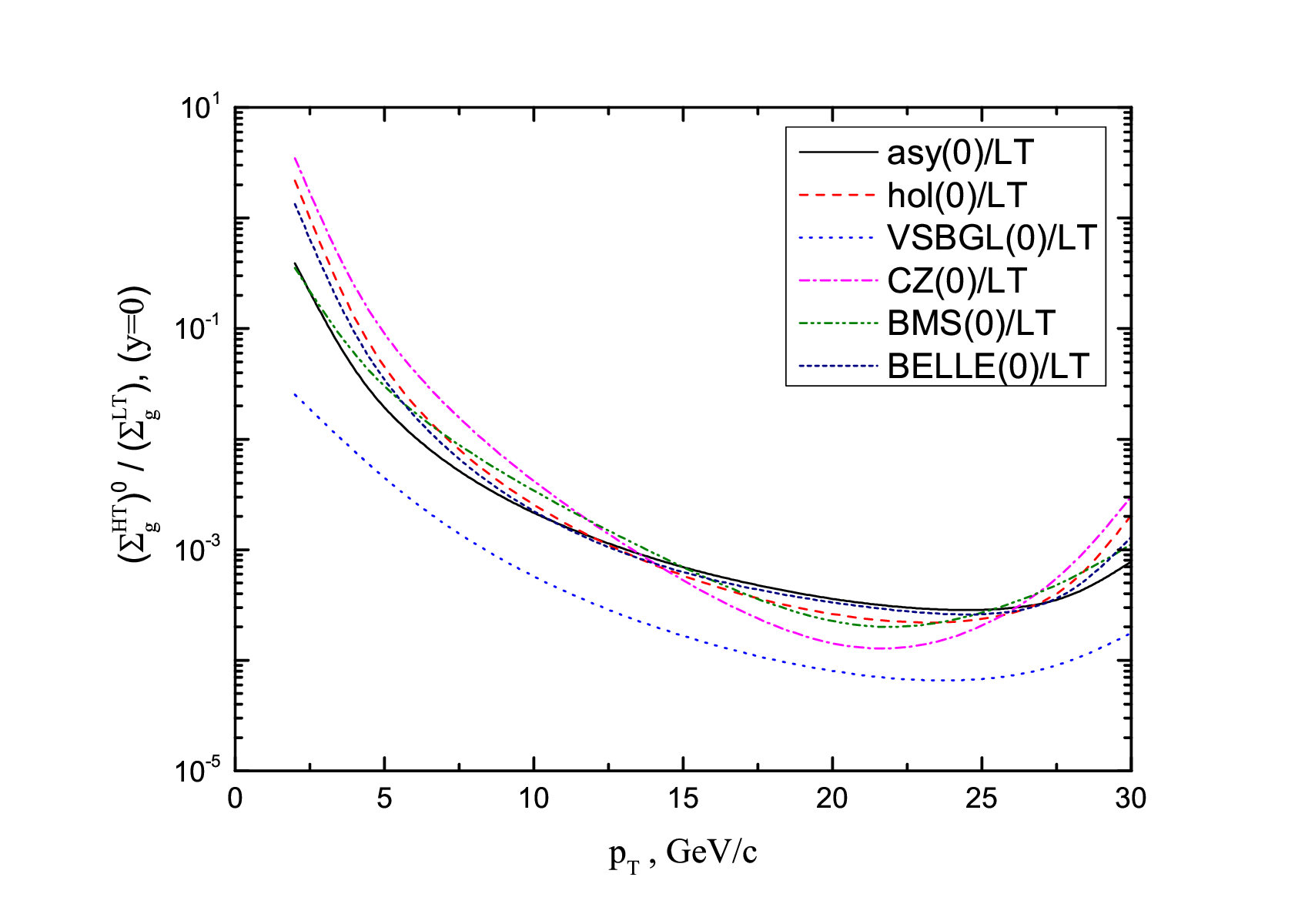}}
\vskip-0.2cm \caption{Ratio
$(\Sigma_{g}^{HT})^{0}/(\Sigma_{g}^{LT})$, in the process $\pi^{-}
p\to g X$, as function of the $p_{T}$ transverse momentum of the
gluon at the c.m. energy $\sqrt s=62.4\,\, GeV$.} \label{Fig10}
\end{figure}

\begin{figure}[!hbt]
\vskip-1.2cm\epsfxsize 11.8cm \centerline{\epsfbox{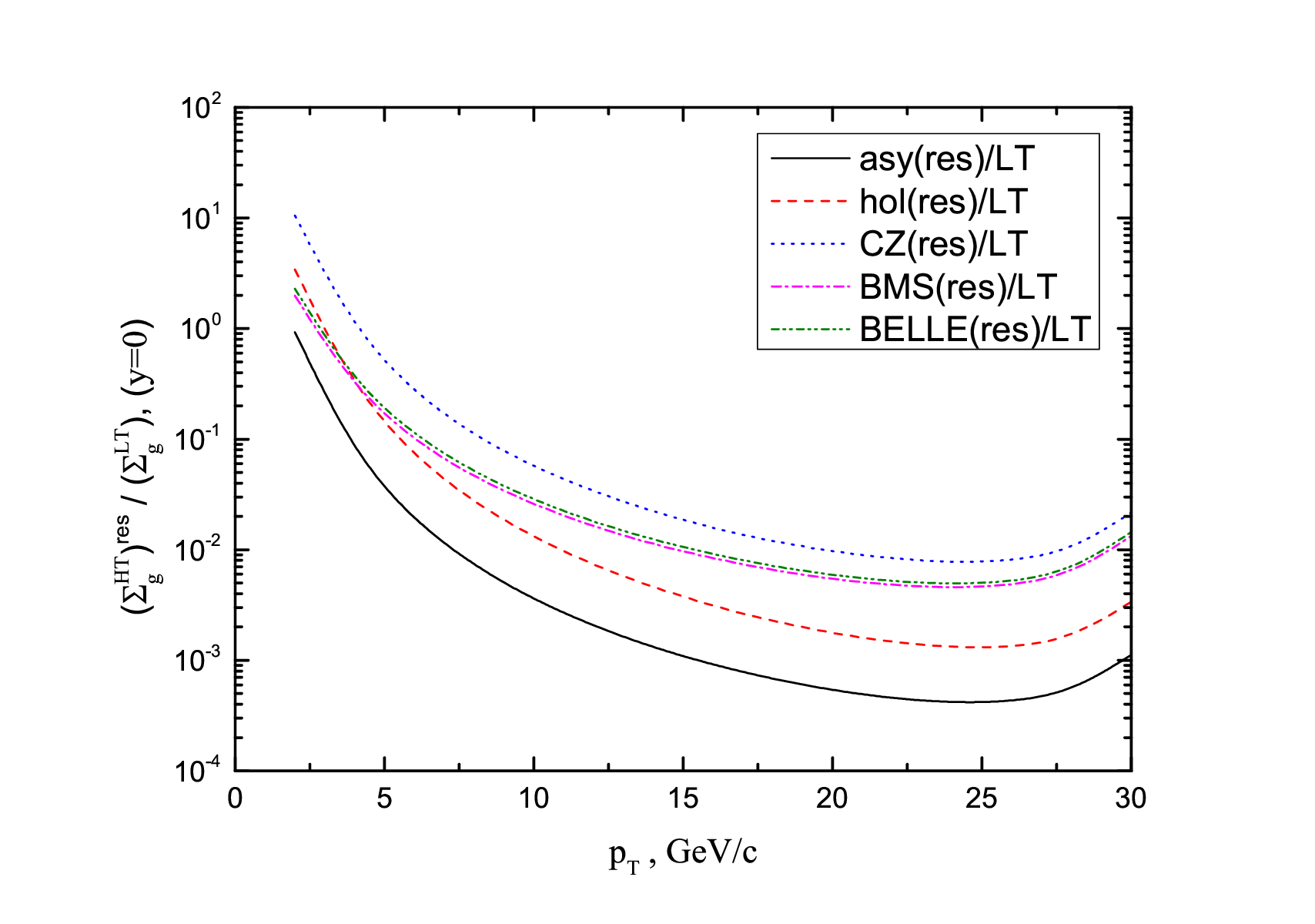}}
\vskip-0.2cm \caption{Ratio
$(\Sigma_{g}^{HT})^{res}/(\Sigma_{g}^{LT})$, in the process $\pi^{-}
p\to g X$, as function of the transverse momentum of the gluon
$p_{T}$ at the c.m. energy $\sqrt s=62.4\,\, GeV$.} \label{Fig11}
\end{figure}

\begin{figure}[!hbt]
\vskip-1.2cm\epsfxsize 11.8cm \centerline{\epsfbox{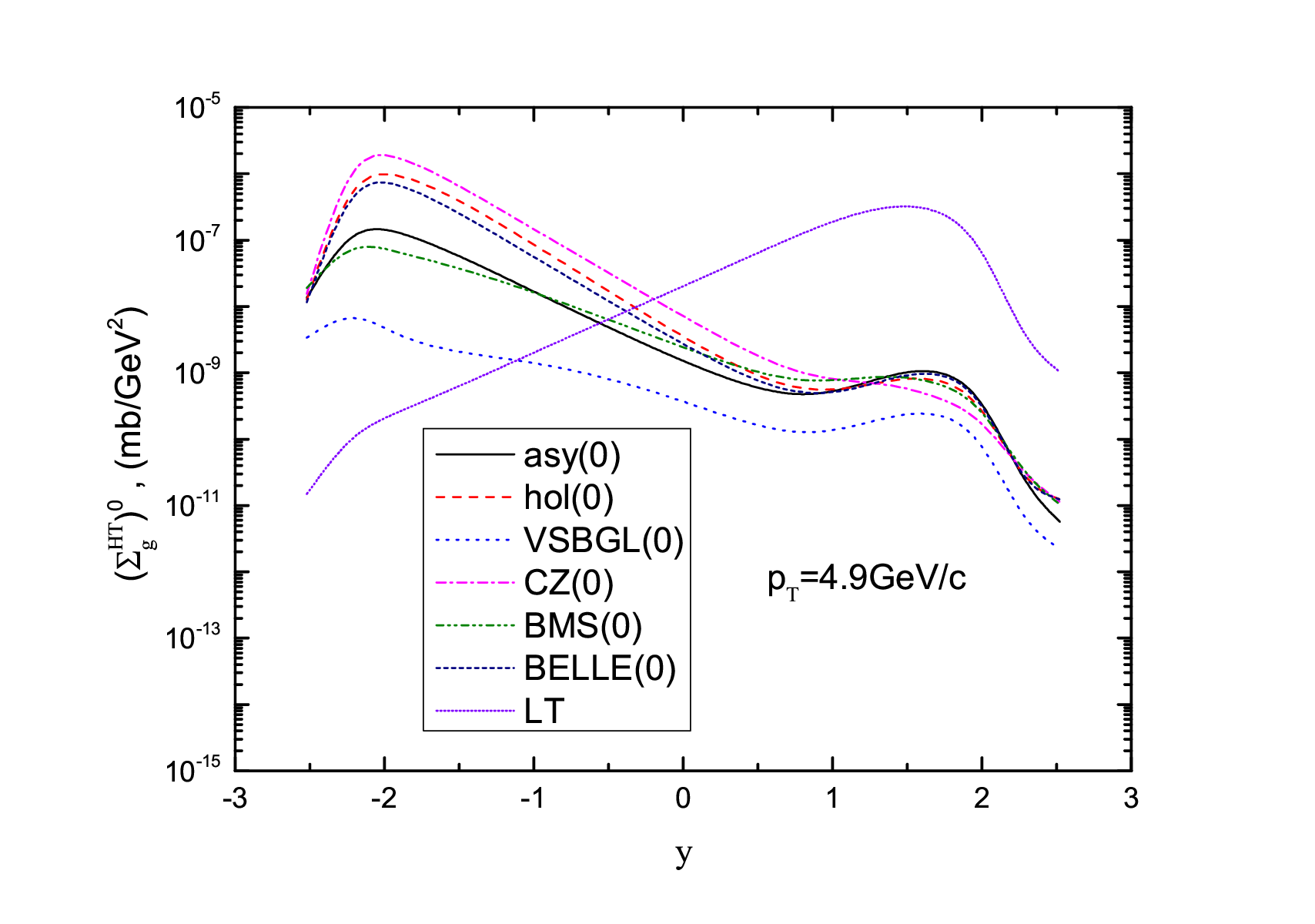}}
\vskip-0.2cm \caption{Higher-twist $\pi^{+} p\to g X$ inclusive
gluon production cross-section $(\Sigma_{g}^{HT})^{0}$, as function
of the rapidity of the gluon  $y$ at the  transverse momentum of the
gluon $p_T=4.9\,\, GeV/c$, at the c.m. energy $\sqrt s=62.4\,\,
GeV$.} \label{Fig12}
\end{figure}

\begin{figure}[!hbt]
\vskip-1.2cm\epsfxsize 11.8cm \centerline{\epsfbox{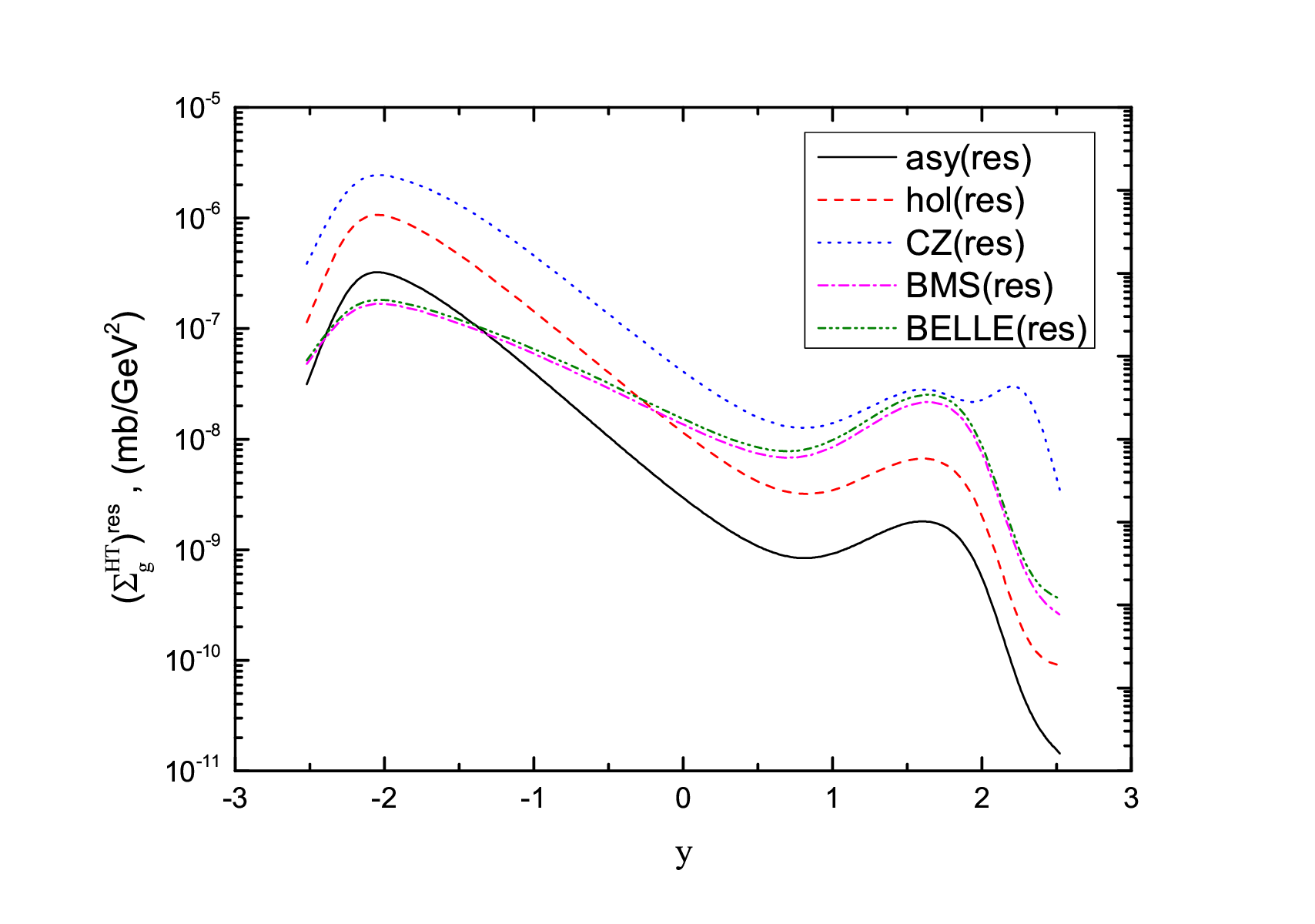}}
\vskip-0.2cm \caption{Higher-twist $\pi^{+} p\to g X$ inclusive
gluon production cross-section $(\Sigma_{g}^{HT})^{res}$, as a
function of the rapidity of the gluon $y$  at the  transverse
momentum of the gluon $p_T=4.9\,\, GeV/c$, at the c.m. energy $\sqrt
s=62.4\,\, GeV$.} \label{Fig13}
\end{figure}

\clearpage

\begin{figure}[!hbt]
\vskip -1.2cm \epsfxsize 11.8cm \centerline{\epsfbox{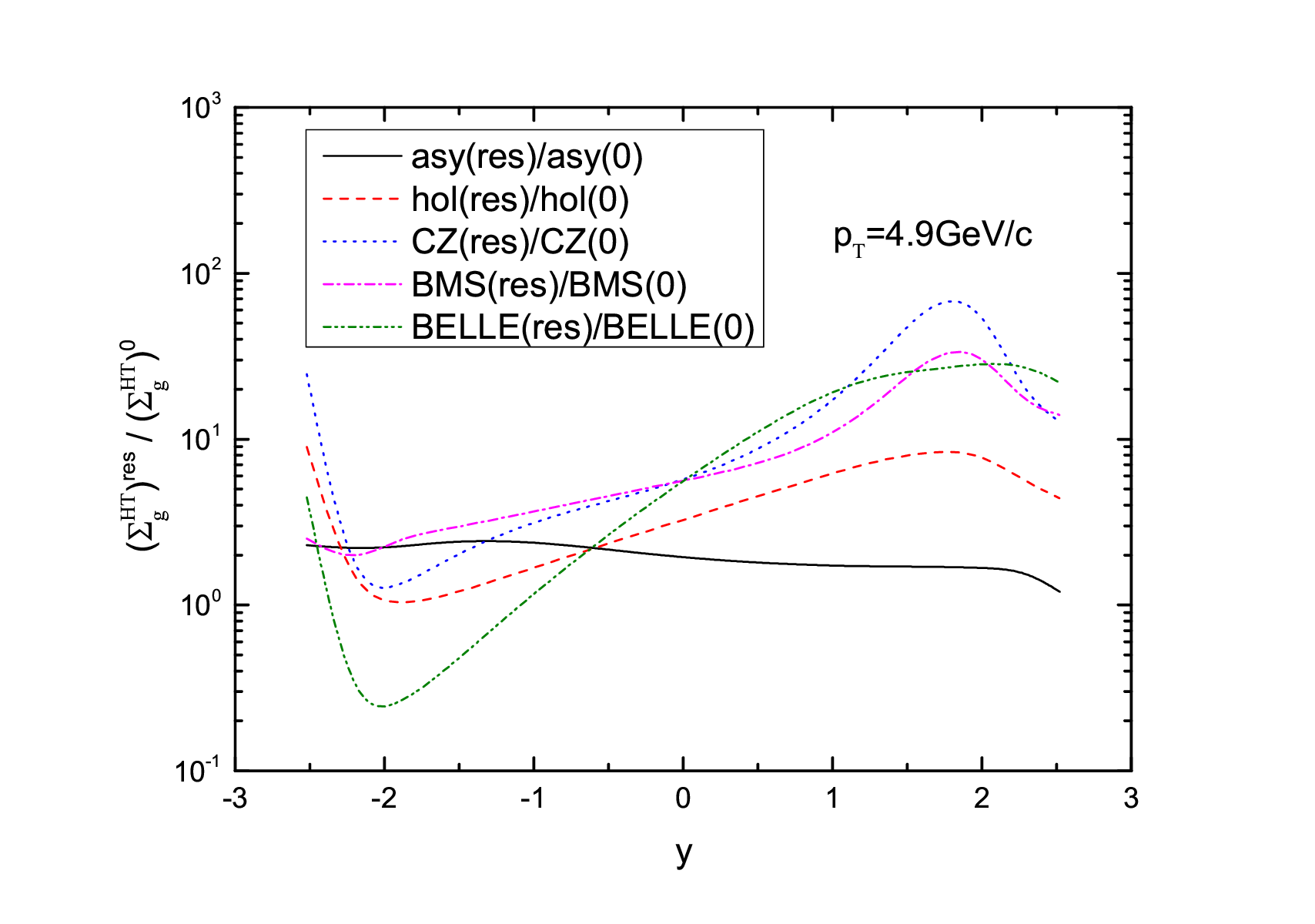}}
\vskip-0.2cm \caption{Ratio
$(\Sigma_{g}^{HT})^{res}/(\Sigma_{g}^{HT})^{0}$, in the process
$\pi^{-} p\to g X$, as function of the rapidity of the gluon $y$ at
the  transverse momentum of the gluon $p_T=4.9\,\, GeV/c$, at the
c.m. energy $\sqrt s=62.4\,\, GeV$.} \label{Fig14}
\end{figure}

\begin{figure}[!hbt]
\vskip -1.2cm \epsfxsize 11.8cm \centerline{\epsfbox{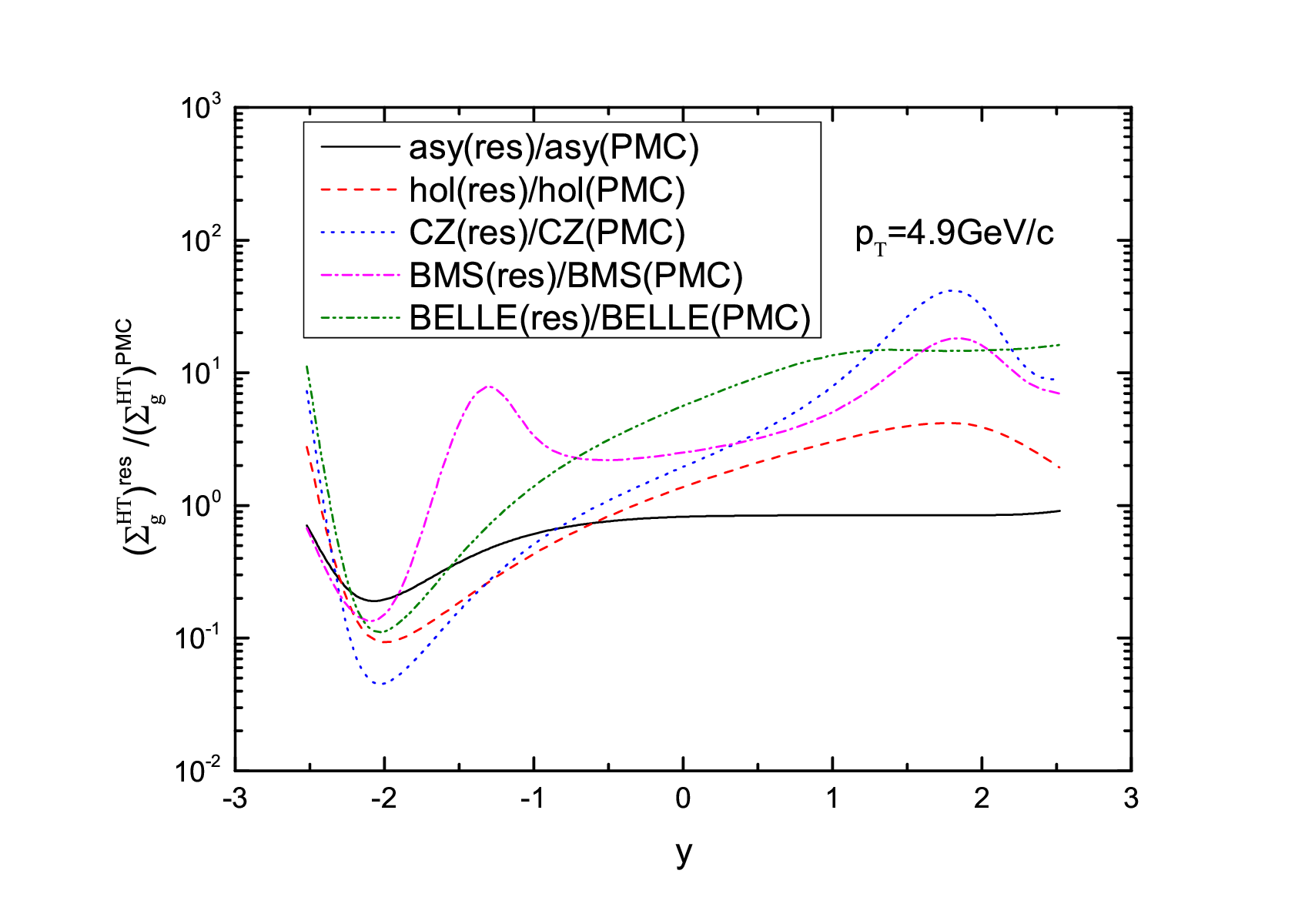}}
\vskip-0.2cm \caption{Ratio
$(\Sigma_{g}^{HT})^{res}/(\Sigma_{g}^{HT})^{PMC}$, in the process
$\pi^{-} p\to g X$, as function of the rapidity of the gluon $y$ at
the  transverse momentum of the gluon $p_T=4.9\,\, GeV/c$, at the
c.m. energy $\sqrt s=62.4\,\, GeV$.} \label{Fig15}
\end{figure}

\begin{figure}[!hbt]
\vskip -1.2cm \epsfxsize 11.8cm \centerline{\epsfbox{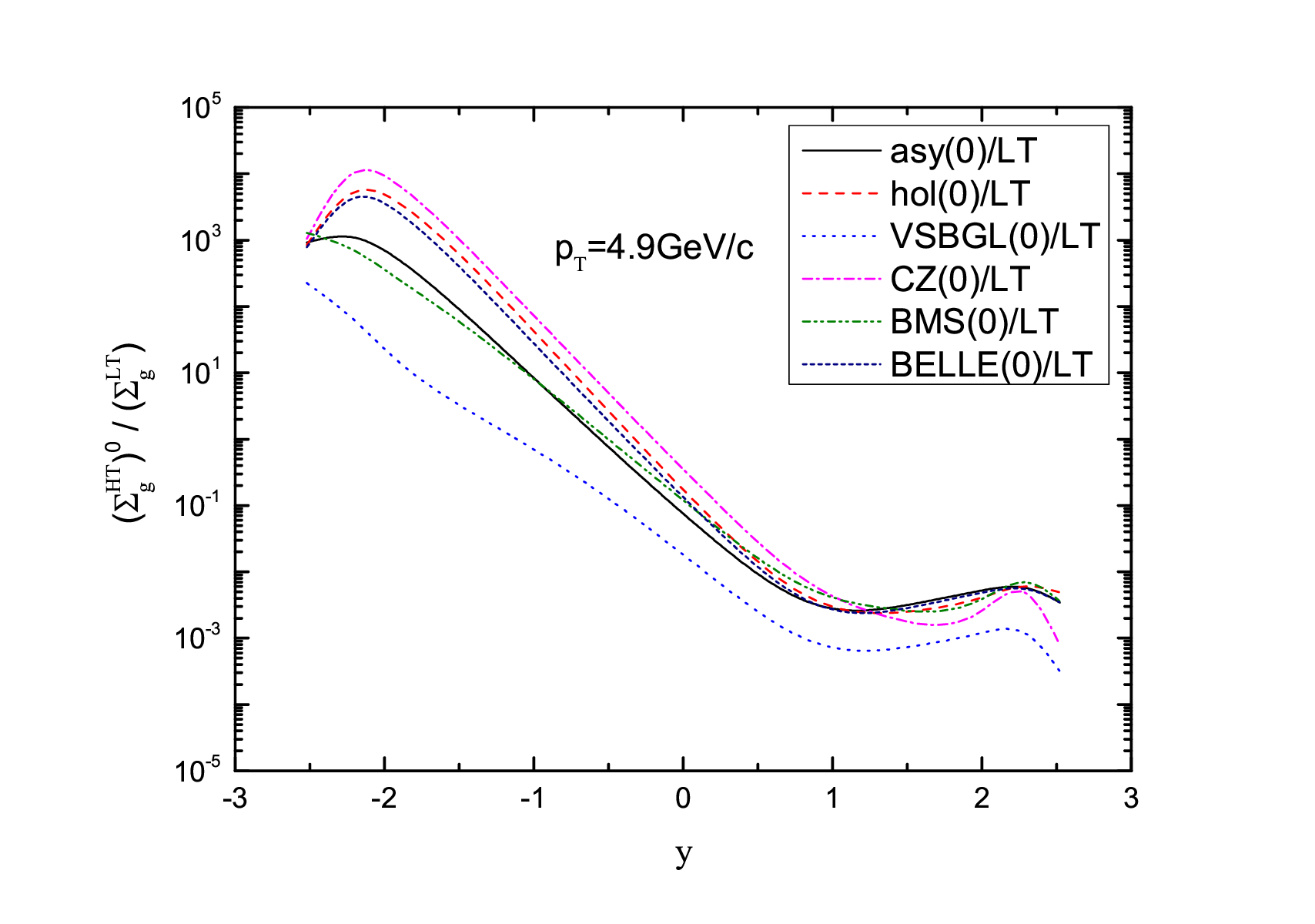}}
\vskip-0.2cm \caption{Ratio
$(\Sigma_{g}^{HT})^{0}/(\Sigma_{g}^{LT})$ in the process $\pi^{+}
p\to g X$, as function of the rapidity of the gluon $y$ at the
transverse momentum of the gluon $p_T=4.9\,\, GeV/c$, at the c.m.
energy $\sqrt s=62.4\,\, GeV$.} \label{Fig16}
\end{figure}

\begin{figure}[!hbt]
\vskip -1.2cm \epsfxsize 11.8cm \centerline{\epsfbox{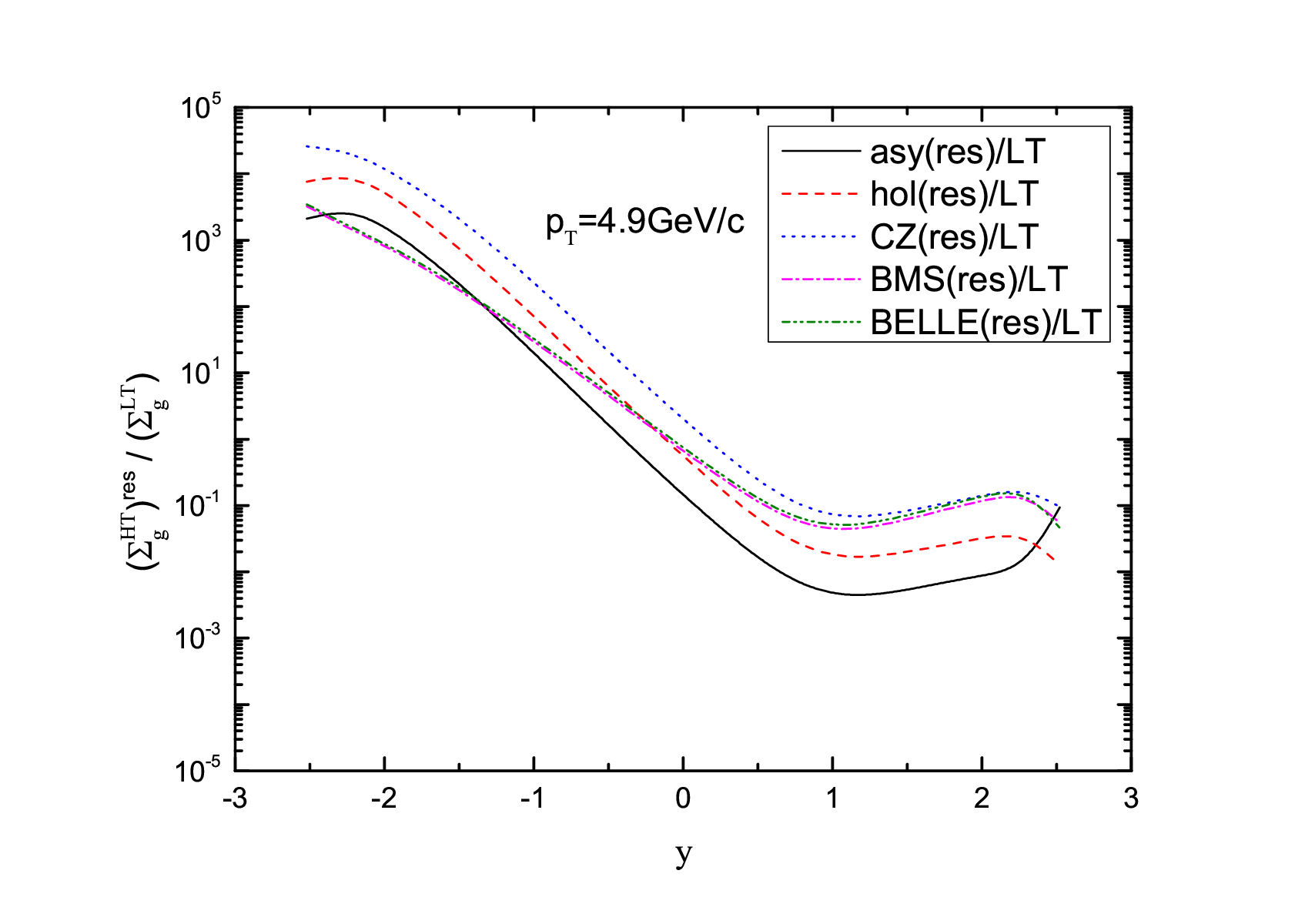}}
\vskip-0.2cm \caption{Ratio
$(\Sigma_{g}^{HT})^{res}/(\Sigma_{g}^{LT})$ in the process $\pi^{+}
p\to g X$, as  function of the rapidity of the gluon $y$ at the
transverse momentum of the gluon $p_T=4.9\,\, GeV/c$, at the c.m.
energy $\sqrt s=62.4\,\, GeV$.} \label{Fig17}
\end{figure}

\begin{figure}[!hbt]
\vskip -1.2cm \epsfxsize 11.8cm \centerline{\epsfbox{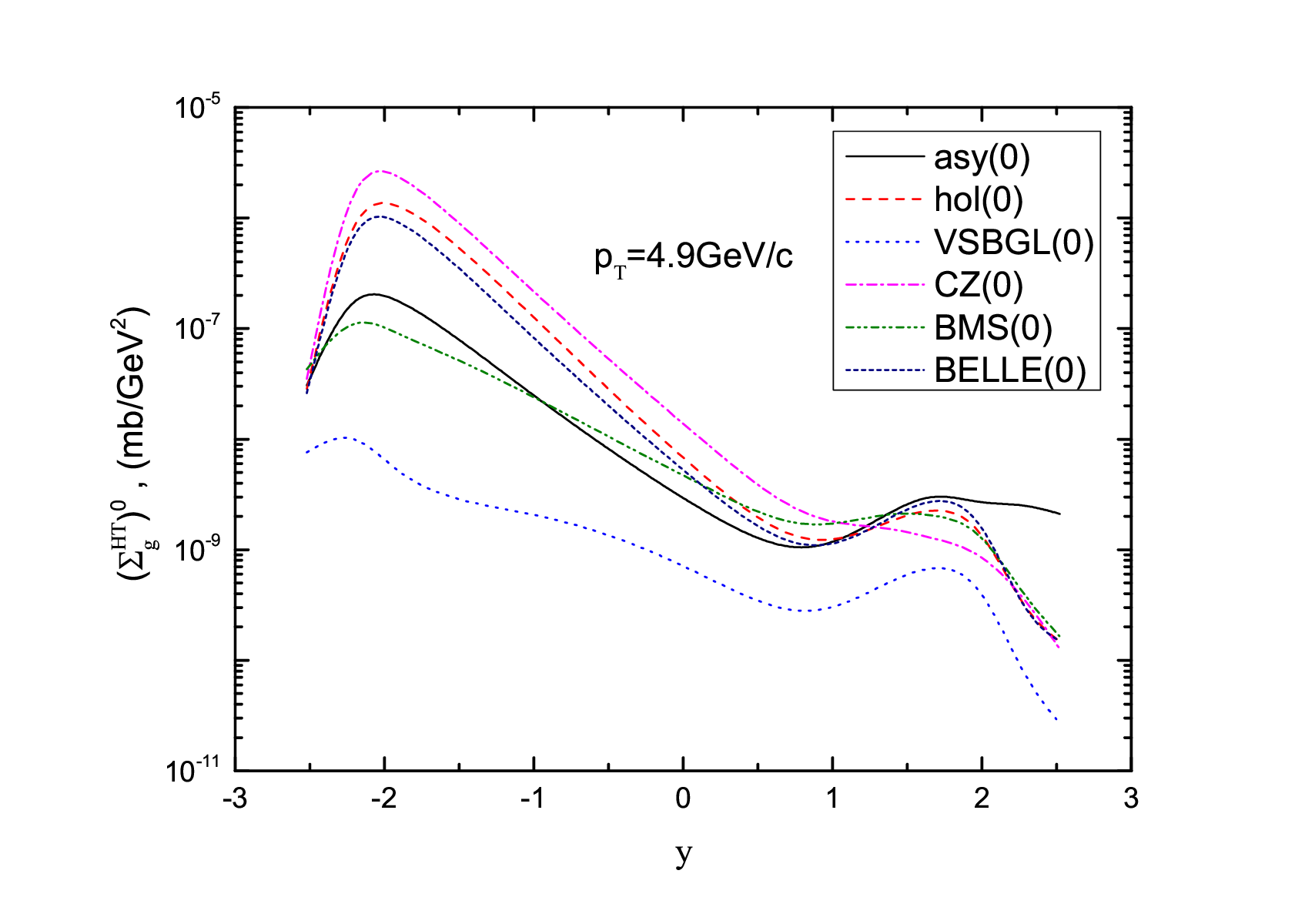}}
\vskip-0.2cm \caption{Higher-twist $\pi^{-} p\to g X$ inclusive
gluon production cross-section $(\Sigma_{g}^{HT})^{0}$, as function
of the rapidity of the gluon $y$  at the  transverse momentum of the
gluon $p_T=4.9\,\, GeV/c$, at the c.m. energy $\sqrt s=62.4\,\,
GeV$.} \label{Fig18}
\end{figure}

\begin{figure}[!hbt]
\vskip -1.2cm \epsfxsize 11.8cm \centerline{\epsfbox{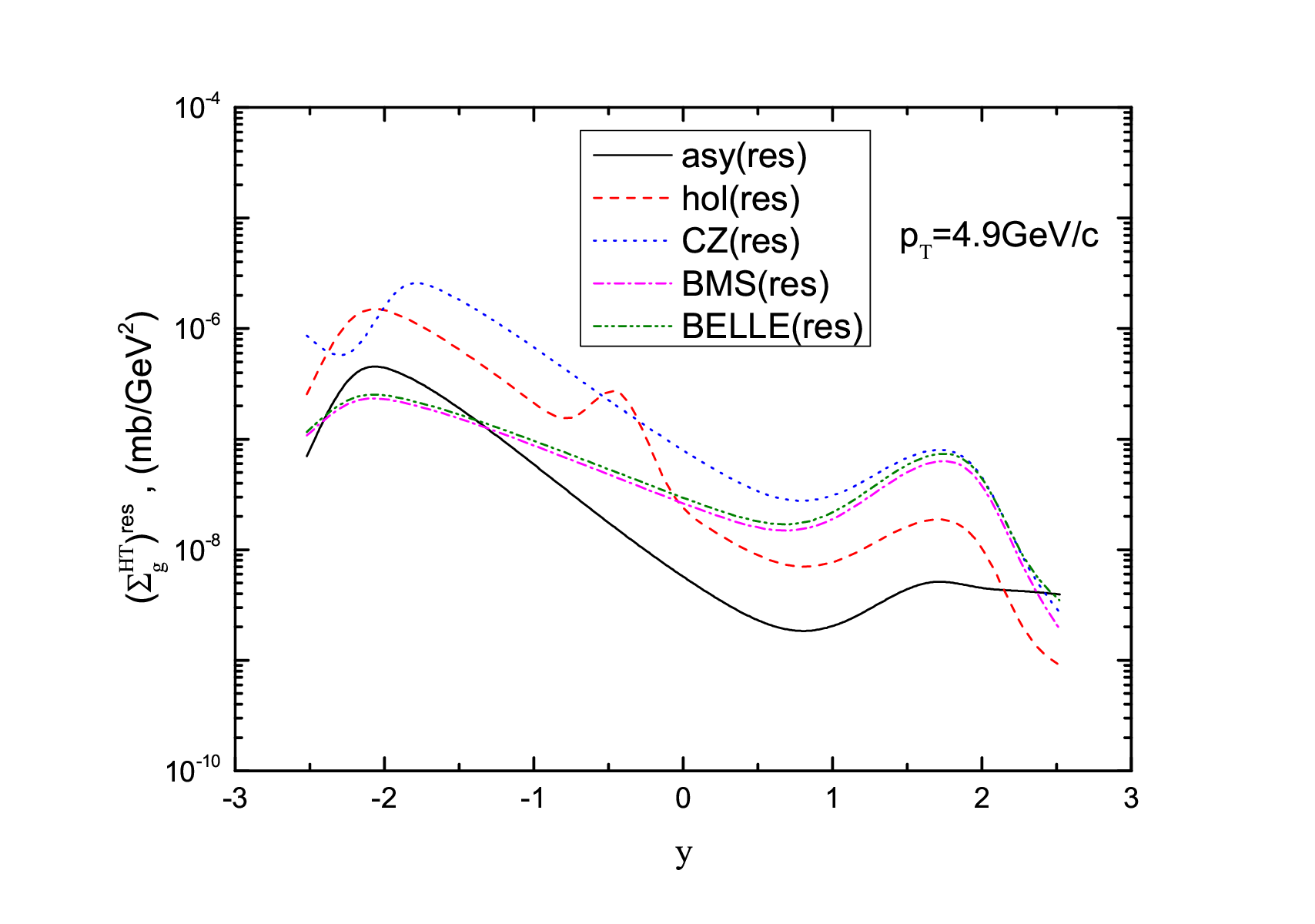}}
\vskip-0.2cm \caption{Higher-twist $\pi^{-} p\to g X$ inclusive
gluon production cross-section $(\Sigma_{g}^{HT})^{res}$, as a
function of the rapidity of the gluon  $y$ at the  transverse
momentum of the gluon $p_T=4.9\,\, GeV/c$, at the c.m. energy $\sqrt
s=62.4\,\, GeV$.} \label{Fig19}
\end{figure}

\begin{figure}[!hbt]
\vskip -1.2cm \epsfxsize 11.8cm \centerline{\epsfbox{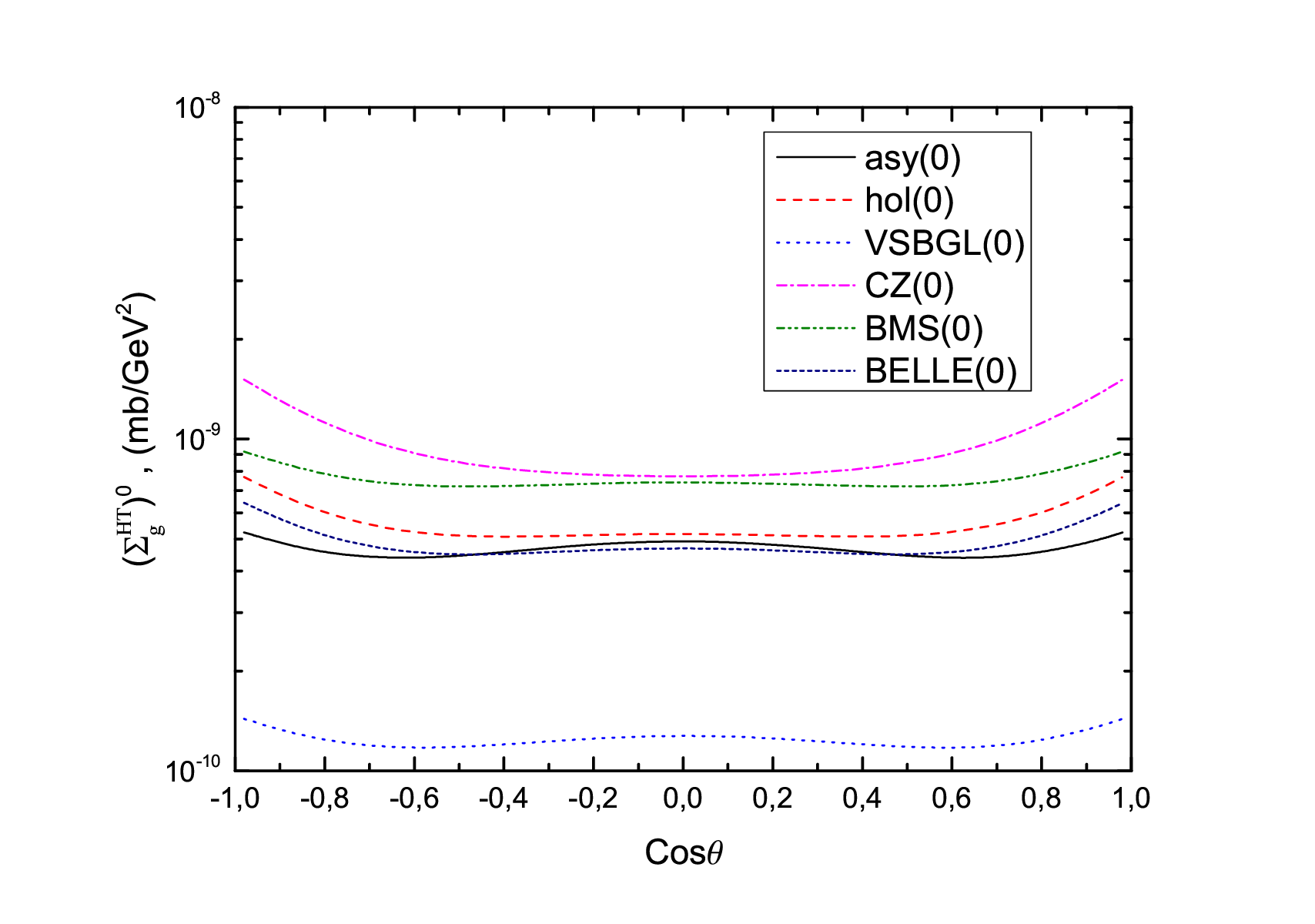}}
\vskip-0.2cm \caption{Angular distributions of higher-twist cross-
sections $(\Sigma_{g}^{HT})^{0}$ for the process $\pi^{+} p\to g X$
at the transverse momentum of the gluon $p_T=4.9\,\, GeV/c$, at the
c.m. energy $\sqrt s=62.4\,\, GeV$.} \label{Fig20}
\end{figure}

\begin{figure}[!hbt]
\vskip -1.2cm \epsfxsize 11.8cm \centerline{\epsfbox{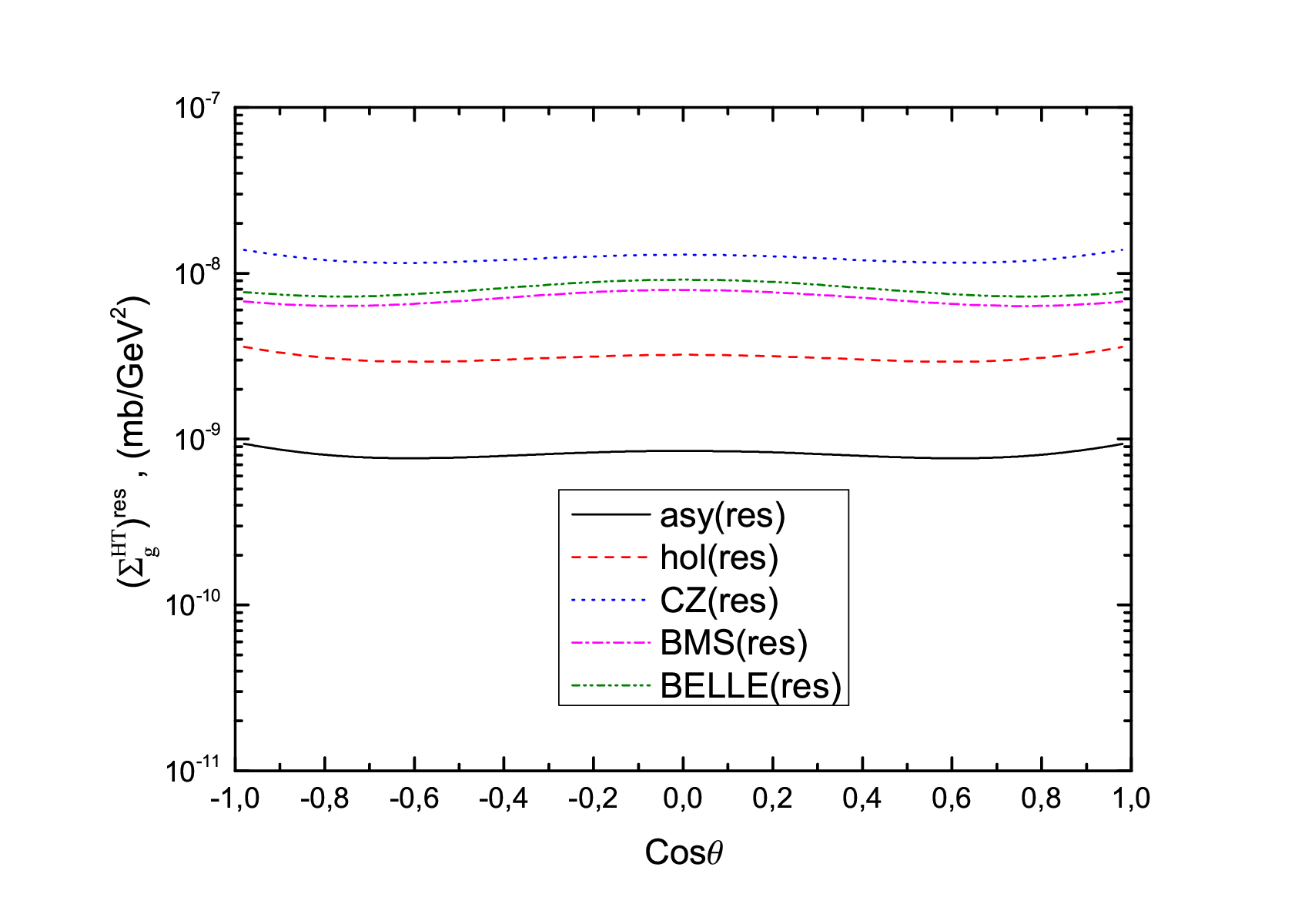}}
\vskip-0.2cm \caption{Angular distributions of higher-twist cross-
sections $(\Sigma_{g}^{HT})^{res}$ for the process $\pi^{+} p\to g
X$, at the transverse momentum of the gluon $p_T=4.9\,\, GeV/c$ at
the c.m. energy $\sqrt s=62.4\,\, GeV$.} \label{Fig21}
\end{figure}

\begin{figure}[!hbt]
\vskip -1.2cm \epsfxsize 11.8cm \centerline{\epsfbox{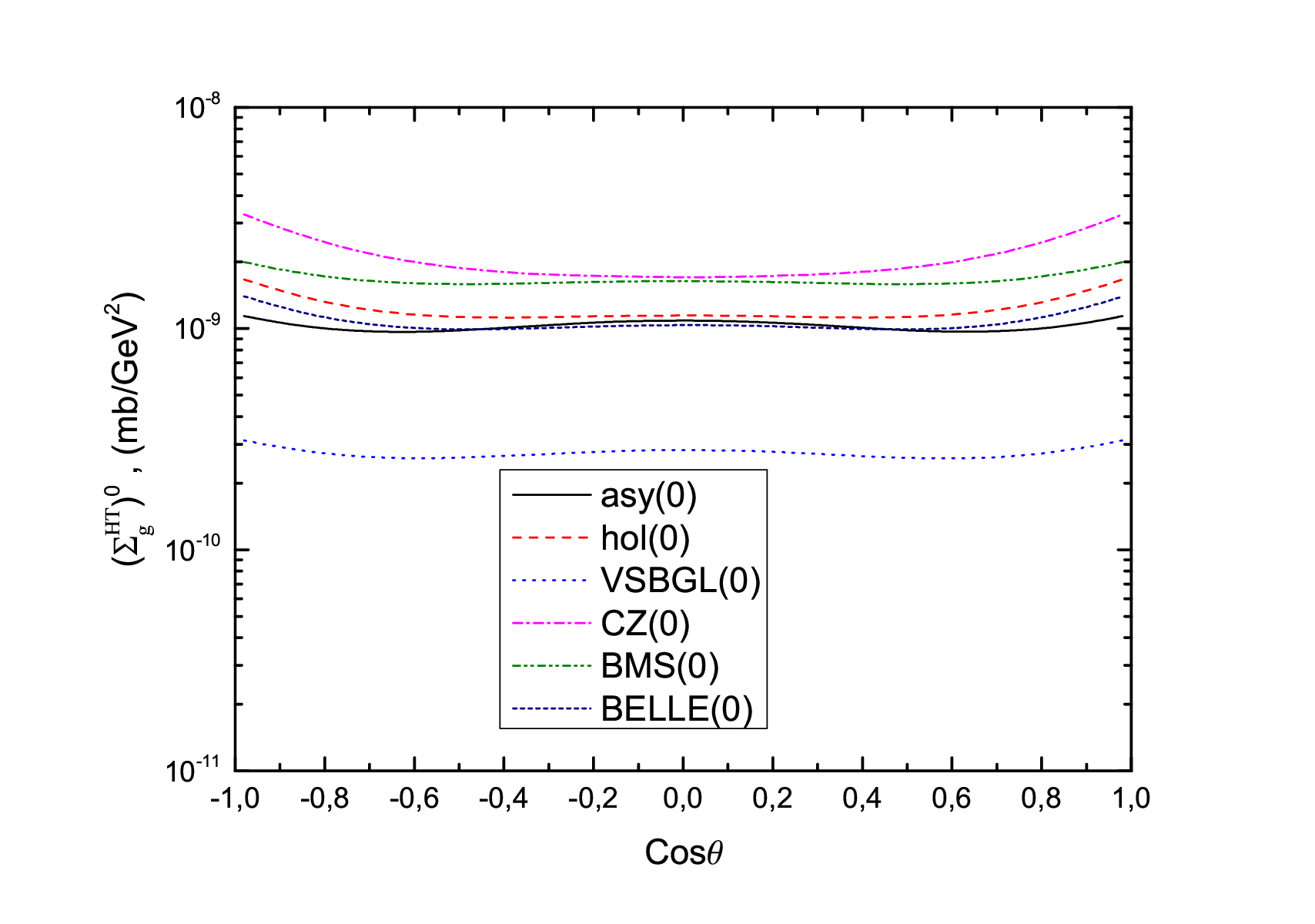}}
\vskip-0.2cm \caption{Angular distribution of higher-twist cross-
sections $(\Sigma_{g}^{HT})^{0}$ for the process $\pi^{-} p\to g X$
at the transverse momentum of the gluon $p_T=4.9\,\, GeV/c$, at the
c.m. energy $\sqrt s=62.4\,\, GeV$.} \label{Fig22}
\end{figure}

\begin{figure}[!hbt]
\vskip 0.8cm \epsfxsize 11.8cm \centerline{\epsfbox{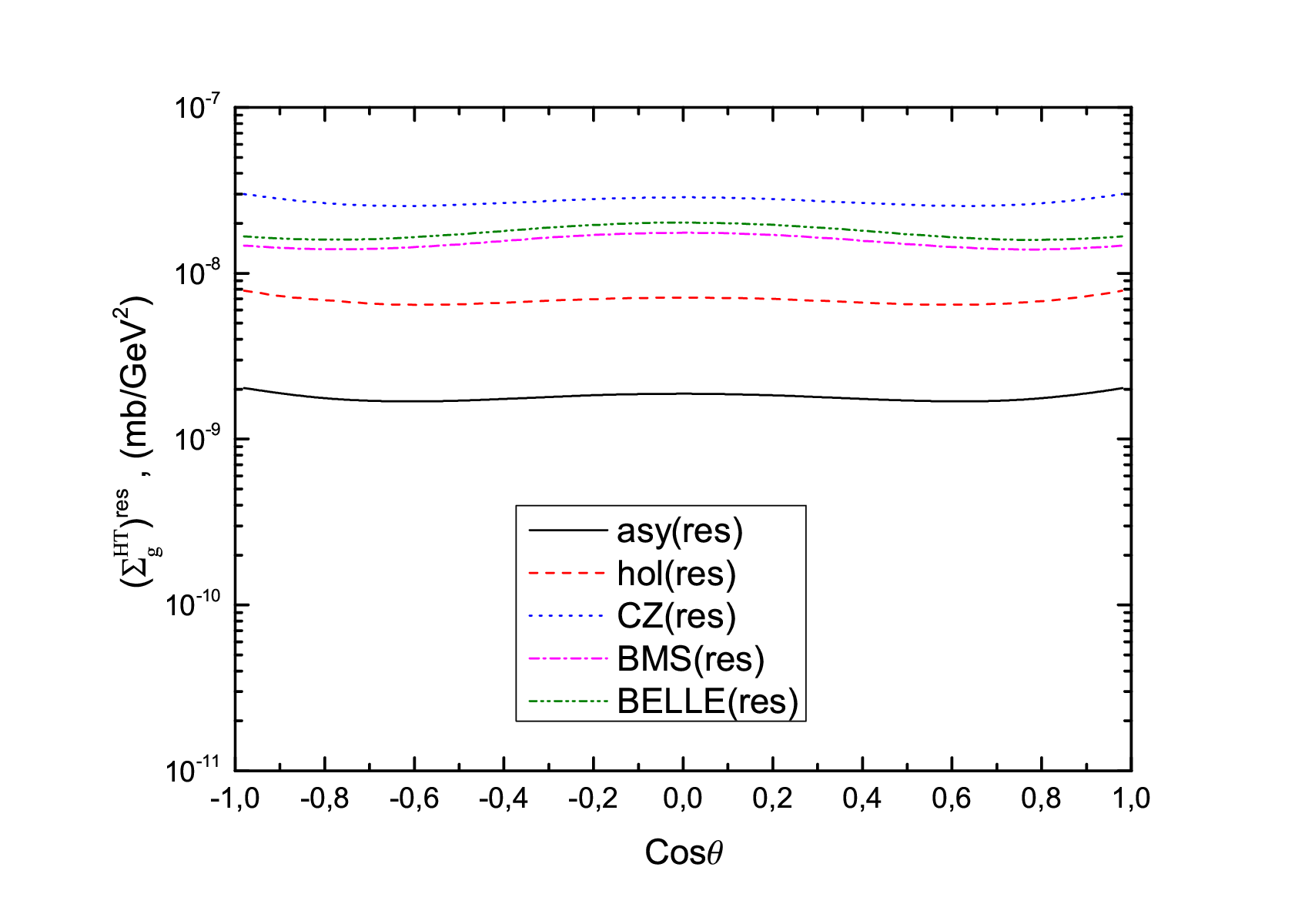}}
\vskip-0.2cm \caption{Angular distributions of  higher-twist cross-
sections $(\Sigma_{g}^{HT})^{res}$ for the process $\pi^{-} p\to g
X$ at the transverse momentum of the gluon $p_T=4.9\,\, GeV/c$, at
the c.m. energy $\sqrt s=62.4\,\, GeV$.} \label{Fig23}
\end{figure}

\end{document}